\documentclass[aps,prb,showpacs,twocolumn,superscriptaddress]{revtex4-2}
\usepackage{amssymb}
\usepackage{graphicx}
\usepackage{appendix}
\usepackage{dcolumn}
\usepackage{bm}
\usepackage{amsmath}
\usepackage{epstopdf}
\usepackage{xcolor}
\usepackage{float}
\usepackage{braket}
\usepackage[colorlinks,urlcolor=blue,anchorcolor=blue,linkcolor=blue,citecolor=blue,breaklinks=true]{hyperref}
\usepackage{multirow}
\usepackage{array}
\usepackage{booktabs}
\usepackage{diagbox}
\usepackage{natbib}

\begin{document}
\title{Hyperbolic fractional Chern insulators}
\author{Ai-Lei He}
\email{heailei@yzu.edu.cn}
\affiliation{College of Physics Science and Technology, Yangzhou University, Yangzhou 225002, China}
\author{Lu Qi}
\affiliation{College of Physics Science and Technology, Yangzhou University, Yangzhou 225002, China}
\author{Yongjun Liu}
\email{yjliu@yzu.edu.cn}
\affiliation{College of Physics Science and Technology, Yangzhou University, Yangzhou 225002, China}
\author{Yi-Fei Wang}
\affiliation{Zhejiang Institute of Photoelectronics $\&$ Zhejiang Institute for Advanced Light Source, Zhejiang Normal University, Jinhua 321004, China}
\affiliation{Center for Statistical and Theoretical Condensed Matter Physics, and Department of Physics, Zhejiang Normal University, Jinhua 321004, China}

\date{\today}

\begin{abstract}
Fractional Chern insulators (FCIs) have attracted intensive attention for the realization of fractional quantum Hall states in the absence of an external magnetic field. Most of FCIs have been proposed on two-dimensional (2D) Euclidean lattice models with various boundary conditions. In this work, we investigate hyperbolic FCIs which are constructed in hyperbolic geometry with constant negative curvature. Through the studies on hyperbolic analogs of kagome lattices with hard-core bosons loaded into topological flat bands, we find convincing numerical evidences of two types of $\nu=1/2$ FCI states, {\emph {i.e.}}, the conventional and unconventional FCIs. Multiple branches of edge excitations and geometry-dependent wave functions for both conventional and unconventional $\nu=1/2$ FCI states are revealed. Intriguingly, the geometric degree of freedom plays various roles for these two FCIs. Additionally, a center-localized orbital plays a crucial role in the unconventional FCI state.
\end{abstract}

\maketitle

\section{Introduction}
Fractional Chern insulators (FCIs)~\cite{Neupert,Tang,Sheng1,YFW1,Regnault1,Qi,GPP,YFW2,Parameswaran,YLWu,WYF3,ZLiu1,Scaffidi,YLWu0,Ronny0,YFW4,ZLiu2,ZLiu3,Ronny1,FCI_reviews,FCI_reviews1,FCI_reviews2}, which realize fractional quantum Hall states without an external magnetic field, have been theoretically proposed for more than a decade. Most recently, there are significantly experimental advances in the realization of FCIs in Moir{\'e} superlattice systems~\cite{Cai2023,Zeng2023,Park2023,XuFan2023,lu2023fractional}. To achieve FCI states, topological flat band (TFB) models are required because TFBs can quench kinetic energy and enhance interaction effectively, which is in analogy of Landau levels (LLs)~\cite{Neupert,Tang,Sheng1,YFW1,Regnault1,FCI_reviews,FCI_reviews1,FCI_reviews2}. By tuning hopping parameters of Chern insulator (CI) models, a series of TFB models have been proposed~\cite{TFBCB,TFBKapit,TFBRuby,TFBStar,TFBdice,TFBKG1,TFBSQOC,Lan}. Due to the similarity between TFBs and LLs, one can define the mapping relationship between single-particle states in CIs and LL wave functions (WFs) in quantum Hall states~\cite{Qi}. Subsequently, several trial WFs for FCIs can be explicitly constructed in Euclidean lattices with various boundary conditions~\cite{Qi,YLWu,YLWu,YLWu0,YHZhang}. Inspired by the analytic expression of Laughlin WFs~\cite{Laughlin1983} and the generalized Pauli principle (GPP)~\cite{GPP1,GPP2,GPP3}, some of us have proposed a direct and effective approach to construct FCI states in disk geometry~\cite{HeAL1,HeAL2} based on the Jack polynomials (Jacks)~\cite{Bernevig1,Bernevig2,Bernevig3} and single-particle states of TFBs. Trial WFs for FCIs in singular CI models have been constructed directly~\cite{HeAL3,HeAL4} as well and the inherent geometric factor for FCIs have been revealed~\cite{HeAL3}. Different from CIs in disk geometry, a defect-core state emerges in singular CI models. Intriguingly, the defect-core orbital in the TFB models can lead to multiple branches of edge excitations (EEs) and two types of $\nu=1/2$ FCIs in which the geometry plays the same role~\cite{HeAL4}.

\begin{figure}[!htp]
\includegraphics[scale=0.75]{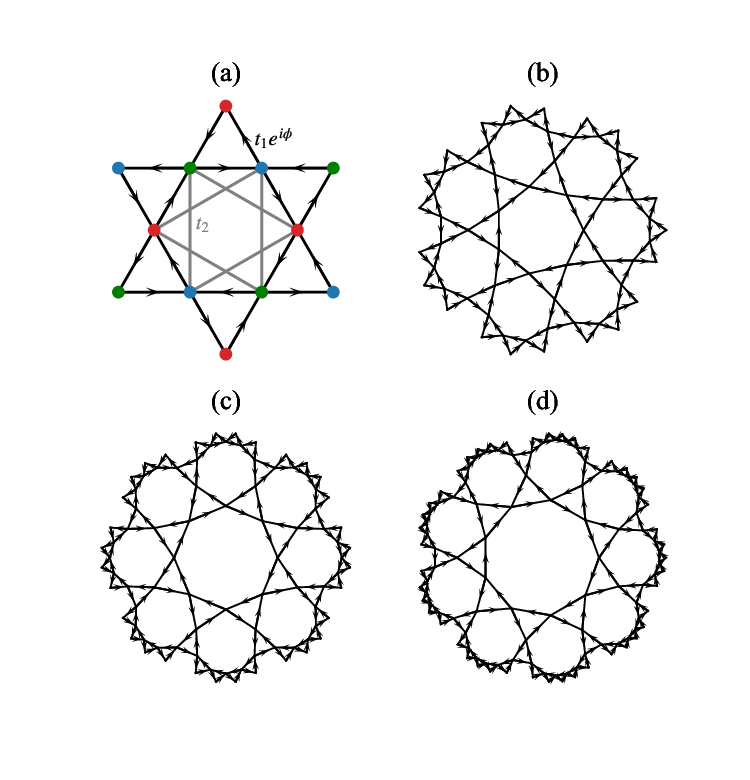}
\caption{(color online). Kagome and kagomelike lattice models. (a) Euclidean kagome lattice model host three atoms in a unit cell which are colored with red, blue and green, respectively. To obtain a nearly TFB, we introduce the NN hopping with staggered magnetic fluxes ($t_1e^{i\phi}$) and the NNN hopping $t_2$. We can introduce the staggered magnetic fluxes in (b) the HKG, (c) the OKG and (d) the NKG lattices. The phase difference are represented by the arrows and these hyperbolic lattices are represented in the Poincar{\'e}-disk model.}
\label{Model}
\end{figure}

Unlike singular surfaces with a point of singular curvature in real space, two-dimensional (2D) hyperbolic geometry has constant negative curvature, but without point singularities. As a new class of 2D lattice structures, hyperbolic lattice models constructed in the hyperbolic plane host several exotic physics phenomena which are beyond in 2D Euclidean lattice models, such as hyperbolic band theory and crystallography~\cite{Joseph,Joseph1,Boettcher,ChengN,Kazuki}, generalized Bloch theory and non-Abelian Bloch states~\cite{ChengN,Elliot,Patrick}, periodic boundary conditions and thermodynamic limit~\cite{Lux,Mosseri,NobleG}, unusual flat band~\cite{FBH1,FBH2} and TFBs~\cite{TFBZhang,Guanpre}, large disorder critical strengths in Anderson localization~\cite{curtis2023,chenA}, effects of strong correlations~\cite{Zhuxc,BieniasP,NobleG2}, Hofstadter spectra~\cite{Kazuki,Kazuki2,Kazuki2,YuS,Stegmaier} and topological states~\cite{YuS,HyperTI11,Zhang2022,LiuZR,Zhang2023,LiuZR1,PeiQS,TaoYL,TarunT,Anffany}, non-Hermitian effect~\cite{SunJ}, holographic duality~\cite{Basteiro}, etc. Experimentally, 2D hyperbolic lattice structures have been realized in circuit quantum electrodynamics (cQED)~\cite{Kollar2019,Boettcher1} and classical electric-circuit networks~\cite{TFBZhang,Lenggenhager2022,Chen2023,Zhang2022,Zhang2023,PeiQS}. Recent theoretical and experimental advances have stimulated interest in the study of condensed matter systems in hyperbolic lattice models. A very challenging topic is whether the existence of FCIs in hyperbolic lattices and how to characterize them.

In this work, we explicitly demonstrate the existence of two types of $\nu=1/2$ bosonic FCI states in hyperbolic analogs of kagome lattice models based on the exact diagonalization (ED) results, {\emph {i.e.}}, the conventional and unconventional FCIs. These two types of FCIs can be characterized based on the EEs and trial WFs for the ground state (GS). According to the exact numerical results, we find more than one branches of EEs and geometry-dependent GS WFs for these $\nu=1/2$ bosonic FCIs. Different from the conventional FCI state, the unconventional FCI state depends on the low-energy center-localized orbital created by the geometry of hyperbolic lattices. With the aid of trial WFs, we find the geometry of hyperbolic lattices plays different roles in these two types of FCIs, in sharp contrast to $\nu=1/2$ FCIs in singular lattices~\cite{HeAL4}.


\section{Models and topological flat bands}
The Euclidean kagome lattice model consists of triangles and hexagons [as illustrated in Fig.~\ref{Model} (a)]. This kagome lattice model hosts a TFB with flatness ratio about 20~\cite{TFBKG1} where the nearest-neighbor (NN) hopping with staggered magnetic fluxes and next-nearest-neighbor (NNN) hopping are considered. The NN and NNN hopping terms are respectively $t_1e^{i\phi}$ and $t_2$, where $\phi$ stems from the staggered magnetic fluxes. In this work, we consider three hyperbolic analogs of kagome lattice models, {\emph {i.e.}}, kagome models made with heptagons, octagons, and nonagons, which are referred to as the heptagon-kagome (HKG), the octagon-kagome (OKG) and the nonagon-kagome (NKG) lattices, respectively. And these hyperbolic lattices are represented using the Poincar{\'e}-disk model [in Fig.~\ref{Model} (b)-(d)]. To obtain TFBs in these kagome-like lattice models, we consider NN and NNN hopping processes similar to the Euclidean kagome TFB model~\cite{TFBKG1}. For the present models, each NN bond carries the phase $\pm\phi$, and the signs of these phases are represented by the directions of arrows [shown in Fig.~\ref{Model} (b)-(d)]. One can easily verify that the total flux of each hyperbolic kagome lattice model is vanished. The tight-binding Hamiltonian of these hyperbolic lattice models can be written as
\begin{eqnarray}
H= &&t_1\sum_{\langle\mathbf{r}\mathbf{r}^{\prime}\rangle}
\left[{\rm e}^{\phi_{\mathbf{r}^{\prime}\mathbf{r}}} a^{\dagger}_{\mathbf{r}^{\prime}}a_{\mathbf{r}}+\textrm{H.c.}\right]+ \nonumber\\ &&t_{2}\sum_{\langle\langle\mathbf{r}\mathbf{r}^{\prime}\rangle\rangle}
\left[a^{\dagger}_{\mathbf{r}^{\prime}}a_{\mathbf{r}}+\textrm{H.c.}\right],
\label{e.1}
\end{eqnarray}
where $a^{\dagger}_{\mathbf{r}}$ ($a_{\mathbf{r}}$) creates (annihilates) a particle at site $\mathbf{r}$. $\langle...\rangle$ and $\langle\langle...\rangle\rangle$ denote the NN and NNN pairs of sites and their corresponding hopping integrals are $t_1$ and $t_2$. $\phi_{\mathbf{r}^{ \prime}\mathbf{r}}=\pm\phi$ is the phase difference between the NN sites.

\begin{figure}[!htp]
\includegraphics[scale=0.65]{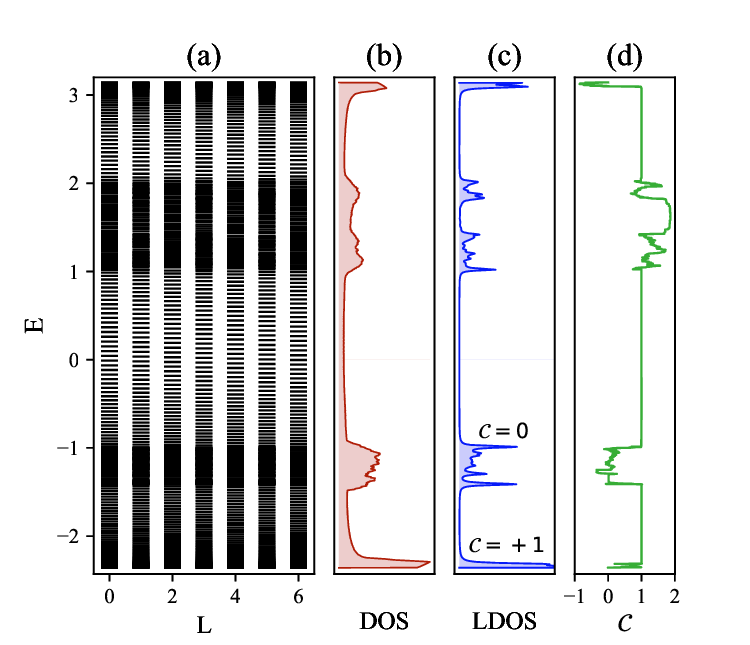}
\caption{(color online). Topological flat ``band" of the HKG lattice model. (a) Energy spectrum of HKG lattice model are arranged in various angular momentum $L$ sectors. (b) the density of states (DOS), (c) the local density of states (LDOS) in the bulk and (d) the real-space Chern number for this energy spectrum. Here, we consider a 2072-site HKG model and the hopping parameters are $t_1=-1.0$, $t_2=0.19$ and $\phi=-0.22\pi$.}
\label{TFBs_HKG}
\end{figure}

Both these hyperbolic kagome lattice models and their corresponding Hamiltonian [Eq.~(\ref{e.1})] respectively host sevenfold, eightfold and ninefold rotational symmetries. One can obtain single-particle energy spectra arranged in various angular momentum sectors because the angular momentum is a good quantum number. To obtain nearly TFBs in the kagome-like lattice models, we adopt the TFB parameters of Euclidean kagome lattice model~\cite{TFBKG1}, {\emph {i.e.}}, $t_1=-1.0$, $t_2=0.19$ and $\phi=-0.22\pi$. Here, we take the HKG model as an example and OKG, NKG lattice models host similar properties [details in the APPENDIX.~\ref{HTFBs}]. The single-particle energy states of the HKG model with open boundary condition is obtained by diagonalizing the Hamiltonian [Eq.~(\ref{e.1})] classified by the quantum number of angular momentum $L=0,1,2,3,...\ {\bf mod} \ 7$  [displayed in Fig.~\ref{TFBs_HKG}(a)]. To characterize features of this energy spectrum, the density of states (DOS), local density of states (LDOS) in the bulk and real-space Chern number are respectively considered in Fig.~\ref{TFBs_HKG}(b)-(d). Based on the DOS and bulk LDOS [ see Fig.~\ref{TFBs_HKG}(b)-(c)], the lowest energy bands seems nearly flat and it has a non-zero real-space Chern number, {\emph {i.e.}}, ${\cal C}=+1$. Consequently, we obtain a TFB in the HKG model. However, this TFB may differ from the TFBs in Euclidean lattice models, because the lowest energy bands of hyperbolic CI models comprise more than one bands with periodic boundary condition on the basis of the hyperbolic band theory and crystallography~\cite{Joseph,Joseph1,Boettcher,ChengN,Kazuki}. The present TFB actually hosts several flat bands with the total Chern number ${\cal C}=1$~\cite{TFBZhang,Guanpre}.

\begin{figure}[!htp]
\includegraphics[scale=0.75]{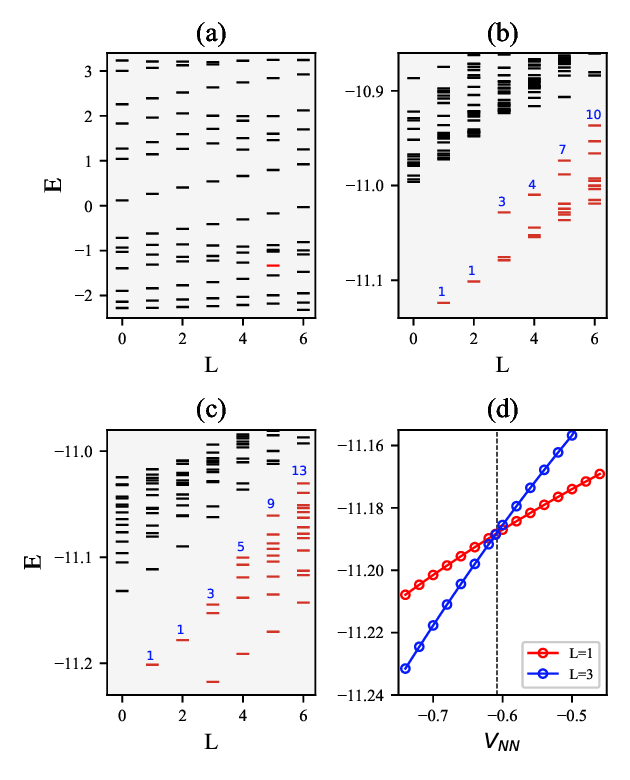}
\caption{(color online). (a) Single-particle energy spectrum of the TFB model in a 98-site HKG lattice. One center-localized state near the lowest band is colored with red. (b) EEs for the conventional $\nu=1/2$ FCIs in the HKG TFB models filling with five hard-core bosons. (c) EEs for the unconventional $\nu=1/2$ FCIs in the HKG TFB models filling with five hard-core bosons. (d) GS energy for $\nu=1/2$ FCIs filling with 5 bosons by tuning the NN interaction potential $V_{\rm NN}$ in HKG TFB model and the crossing point ($V_{\rm NN}\approx-0.61$) is marked by a black dashed line. $L$ is the angular momentum related to the rotational symmetry. The trap potential is $V_{\rm trap}=0.005$.}
\label{EE}
\end{figure}

\section{Edge excitations}
For the FCIs in hyperbolic disk without trap potentials ({\emph i.e.}, the infinite flat potential well), clear edge modes do not exist along the boundary~\cite{JEMoore} (we also present related results in the APPENDIX~\ref{EEs1}). In order to confine the FCI droplet and make the edge modes propagate around the boundary, an additional trap potential is required on the finite-size hyperbolic disks. Here, we choose the  harmonic trap with the form $V=V_{{\rm trap}}\sum_{\bf r} |{\bf r}|^2 n_{\bf r}$~\cite{YFW4,HeAL1,HeAL2,HeAL3,HeAL4,JEMoore}, where $V_{\rm trap}$ is the trap potential strength and $n_{\bf r}$ is the number of particles. $|{\bf r}|$ is the radius from the disk center. The hyperbolic distance between two points $z$, $z^{\prime}$ ($z = x + {\rm i}y$ as the complex coordinate in the Poincar{\'e} disk) is defined as
\begin{eqnarray}
d(z,z^{\prime}) = \kappa {\rm arcosh} {\Bigg(} 1+  \frac{2|z-z^{\prime}|^2}{(1-|z|^2)(1-|z^{\prime}|^2)} {\Bigg)},
\label{e.2}
\end{eqnarray}
where $\kappa$ is related to the corresponding negative curvature $K$, ${\emph i.e.}, \kappa = {\sqrt{-K}}$ and we set $\kappa=1$. Accordingly, $|{\bf r}|={\rm arcosh}{( (1+|z|^2)/(1-|z|^2))}$. The single-particle spectrum of a 98-site HKG model with trap potential $V_{\rm trap}=0.02$ is shown in Fig.~\ref{EE} (a) and the angular momentum of GS is six. Different from single-particle spectrum in 2D Euclidean model, a center-localized state appears near the low-energy band with angular momentum five and the corresponding state is localized around the center of hyperbolic disk [details shown in Fig.~\ref{EE} (a) and the APPENDIX~\ref{EEs1}]. The center-localized state also appears in the OKG and NKG TFB models (see APPENDIX~\ref{EEs2}).

Here, we adopt the real-space ED method to numerically
diagonalize the full many-body Hamiltonian without any band projection. We present the EE spectra for FCIs in the HKG TFB models filling with five hard-core bosons. When considering particles without NN interaction, the total angular momentum of the GS is one and there is a branch of EEs with degeneracy sequences ``1,1,3,4,7,10,\ ..." according to the ED results [as shown in Fig.~\ref{EE} (b)].  Although the degeneracy sequences of the EEs are not ``1,1,2,3,5,7,...", this FCI state seems to be a $\nu=1/2$ FCI for two reasons: $\rm {\romannumeral 1})$ the total angular momentum of $\nu=1/2$ FCI GS is $(6+1+3+5+0) \ \bf mod$ 7=1 (the corresponding root configuration $|1010101010\rangle_{\rm FCI}$ and the occupying configuration shown in the APPENDIX~\ref{EEs1}); $\rm {\romannumeral 2})$ two branches of EEs mix together, and the degeneracy sequences of one branch is ``1,1,2,3,5,7,\ ..." and the other is ``0,0,1,1,2,3,\ ...", which is reminiscent of FCIs in singular lattices where defect-core orbital leads to multiple branches of EEs~\cite{HeAL3,HeAL4}.  The emergence of two branches of EEs for this FCI may be related to the center-localized orbital.

For the FCIs in singular lattices, energy crossing occurs between the GS and the first excited state by tuning the NN interaction and a FCI state with one particle occupied the defect-core orbital appears with attractive interaction~\cite{HeAL4}. Inspired by this, we add NN attractive interaction (we choose $V_{\rm NN}=-0.7$) in the HKG TFB model. Based on the ED results, we obtain a many-body state which degeneracy sequences of the EEs become ``1,1,3,5,9,13,\ ..." and the GS angular momentum is three [as  illustrated in Fig.~\ref{EE} (c)]. This state possibly belongs to a $\nu=1/2$ FCI in which one hard-core boson fills into the center-localized orbital, namely, the unconventional $\nu=1/2$ FCI. The corresponding root configuration may be $|\underline{1}01010101\rangle_{\rm FCI}$ (the occupying configuration in the APPENDIX~\ref{EEs1}), where $\underline{1}$ denotes one particle occupying the center-localized orbital, and the total angular momentum is $(5+0+2+4+6) \ {\bf mod} \ 7 = 3$ consistent with the numerical results. Multiple branches of EEs mix together as well and each branch hosts the degeneracy sequences ``1,1,2,3,5,7,\ ...", ({\emph {i.e.}} ``1,1,2,3,5,7,\ ..." + ``0,0,1,1,2,3,\ ..." + ``0,0,0,1,1,2,\ ..." + ``0,0,0,0,0,1,\ ..."). By tuning the NN interaction potential $V$, energy crossing occurs between the GS and the first excited state [displayed in Fig.~\ref{EE} (d)], similar to the FCIs in singular lattices~\cite{HeAL4}. We find similar results appear in the OKG and NKG TFB models as well (details shown in the APPENDIX~\ref{EEs2} and~\ref{EEs3}). However, whether the geometry in these hyperbolic TFB models plays a role and whether it plays the same role for the conventional and unconventional $\nu=1/2$ FCIs remains unclear.

\section{ Trial wave functions}
Based on the GPP, the Jacks and the single-particle states of TFBs ($\{|\phi_i\rangle\}$), trial WFs for the $\nu=1/m$ FCI in disk or singular geometries can be directly constructed with a general expression, $i.e.$ $\Psi^{\nu=1/m}_{\rm FCI}=\sum_l J_{\lambda_l} \Phi^{\rm TFB}_{\lambda_l}$~\cite{HeAL1,HeAL3} (details also see the APPENDIX~\ref{WF0}), where $\Phi^{\rm TFB}_{\lambda_l}$ is the antisymmetric Slater determinant (for fermions) or symmetric polynomial (for bosons) composed by the single-particle states of TFBs and $J_{\lambda_l}$ is the expansion coefficients which are related to the $l-$th basis configurations $\lambda_l$ which  comes from squeezing the root configuration~\cite{Bernevig1,Bernevig2,Bernevig3}. Here we take $\nu=1/2$ bosonic FCI filling with three bosons as an example. The root configuration is $ \Phi^{\rm TFB}_{[4,2,0]} = |101010\rangle_{\rm TFB}$~\cite{Note}, a symmetric polynomial composed of $|\phi_0\rangle,\ |\phi_2\rangle$ and $|\phi_4\rangle$.  Based on the squeezing rules~\cite{Bernevig1,Bernevig2,Bernevig3}, other basis configurations are obtained, {\emph {i.e.}}, $ \Phi^{\rm TFB}_{[4,1,1]} = |020010\rangle_{\rm TFB}$, $ \Phi^{\rm TFB}_{[3,2,1]} = |011100\rangle_{\rm TFB}$, $ \Phi^{\rm TFB}_{3,3,0} = |100200\rangle_{\rm TFB}$ and $ \Phi^{\rm TFB}_{[2,2,2]} = |003000\rangle_{\rm TFB}$. Accordingly, the GS of $\nu=1/2$ FCI can be directly written as $\Psi^{\nu=1/2}_{\rm FCI}= J_{[4,2,0]} \Phi^{\rm TFB}_{[4,2,0]} + J_{[4,1,1]} \Phi^{\rm TFB}_{[4,1,1]} + J_{[3,2,1]} \Phi^{\rm TFB}_{[3,2,1]} + J_{[3,3,0]} \Phi^{\rm TFB}_{[3,3,0]} + J_{[2,2,2]} \Phi^{\rm TFB}_{[2,2,2]}$. For various geometries of lattice systems, the geometric factor $\beta$ is implicit in the expansion coefficients, $J_{\lambda_l}\equiv J_{\lambda_l}(\beta)$. For example, the geometric factor $\beta=6/n$ is related to the $n$-fold rotational symmetry in singular Kagome lattices and the FCI states are dependent on this geometric factor~\cite{HeAL3,HeAL4}.

\begin{figure}[!htp]
\includegraphics[scale=0.75]{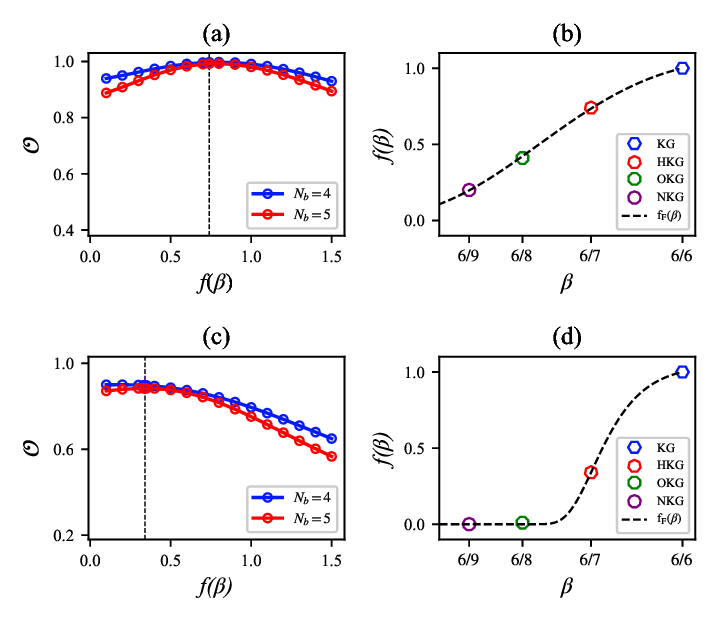}
\caption{(color online). (a) WF Overlap between the trial GS WFs and the ED results with variable geometric function $f(\beta)$ for $\nu=1/2$ FCI filling with $N_b=4$ and $N_b=5$ hard-core bosons in HKG TFB models. (b) Based on the WF Overlaps, the suitable geometric function $f(\beta)$ can be verified for $\nu=1/2$ FCIs in the  Euclidean kagome (KG), HKG, OKG and NKG TFB models, and the fitting function $f_{\rm F}(\beta)$ is obtained. (c) WF overlap for unconventional $\nu=1/2$ FCI filling with $N_b=4$ and $N_b=5$ hard-core bosons in HKG TFB models. (d) The suitable geometric function $f(\beta)$ for unconventional $\nu=1/2$ FCIs in hyperbolic lattices and the fitting function $f_{\rm F}(\beta)$ are obtained.}
\label{overlap}
\end{figure}

Inspirited by the trial WFs for FCIs in singular Kagome lattices, the trial WFs for $\nu=1/2$ FCI in HKG lattice (with sevenfold rotational symmetry, {\emph {i.e.}}, $n=7$) can be written as $\Psi^{\nu=1/m}_{\rm FCI}=\sum_{l} J_{\lambda_l} (f(\beta=6/7)) \Phi^{\rm TFB}_{\lambda_l}$ where $f(\beta)$ is a function which is related to the geometric factor $\beta$. To obtain the expression of $f(\beta)$,  a series of trial WFs for $\nu=1/2$ FCI by tuning $f(\beta)$ from 0.1 to 1.5 are constructed. When $f(\beta=6/7)\approx0.74$, the WF overlap value ({\emph {i.e.}}, ${\cal O}=|\langle \Psi^{\nu=1/2}_{\rm FCI} | \Psi_{\rm ED} \rangle |$, the inner product of trial WF and WF from ED result) is very close to the maximum [details shown in Fig.~\ref{overlap} (a)]. For the conventional $\nu=1/2$ FCI state in the OKG and NKG TFB models, when the geometric functions are respectively $f(\beta=6/8)\approx0.41$ and $f(\beta=6/9)\approx0.20$, the corresponding WF overlaps are close to the maximum ( as shown in the APPENDIX~\ref{WF1}). Based on these numerical results, a fitting function is roughly obtained $f_{\rm F}(\beta=6/n)=\beta^{6/\beta-5}=(6/n)^{n-5}$ for the conventional $\nu=1/2$ FCI in hyperbolic lattices with $n$-fold rotational symmetry [in Fig.~\ref{overlap} (b)].

We next consider the unconventional $\nu=1/2$ FCI which one hard-core boson fills into the center-localized orbital. To obtain this FCI, an attractive interaction $V=-0.7$ is added to the HKG model. Based on the GPP, the Jacks and the center-localized orbital, we construct a series of trial WFs for this $\nu=1/2$ FCI by tuning $f(\beta)$ from 0.1 to 1.5. The WF overlap values are shown in Fig.~\ref{overlap} (c) and the geometric function is about 0.34 ({\emph {i.e.}}, $f(\beta=6/7)\approx0.34$) in terms of the max overlap value. The max WF overlap values is not very high ($\sim$0.897 for four bosons and $\sim$0.883 for five bosons), because the occupied center-localized orbital [colored with red in Fig.~\ref{EE} (a)] is far from the TFB orbitals and even near the high-energy bulk orbital. Trial WFs for this unconventional $\nu=1/2$ FCI in  the OKG and NKG TFB models are constructed as well and their corresponding geometric functions are close to zero (see the APPENDIX~\ref{WF1}). A possible fitting function for the geometric function can be roughly obtained  $f_{\rm F}(\beta=6/n)= \beta^{{(6/\beta)^{(6/\beta-6)}}}=(6/n)^{n^{(n-6)}}$ [in Fig.~\ref{overlap} (d)]. One can clearly find for the conventional and unconventional $\nu=1/2$ FCIs, geometry of hyperbolic lattices plays different roles, in stark contrast to FCIs in singular lattices in which the geometry plays the same role~\cite{HeAL4}.

\section{Summary and discussion}
We explicitly investigate $\nu=1/2$ FCIs in hyperbolic TFB models filling with hard-core bosons, and find the conventional and unconventional FCIs emerge. These two types of FCIs can be identified on the basis of the EEs and the trial WFs. According to the ED results, these FCIs host more than one branches of EEs and the degeneracy sequence of each branch is ``1, 1, 2, 3, 5 ...", which can reveal the characterisation of $\nu=1/2$ FCIs. Based on the GPP, the Jacks and single-particle states, we construct a series of trial WFs and the high WF overlap values manifest that for the conventional FCI state, all particles occupy the TFB orbitals, while for the unconventional FCI state, one particle occupies the center-localized orbital. More intriguingly, we find these FCIs are related to the geometry of hyperbolic lattices and the geometry plays various roles in the conventional and unconventional FCIs. It will be a challenging and significant issue to explore how the geometry plays various roles in future studies.

Our findings might open up several future directions on hyperbolic FCIs. The center-localized orbital plays a crucial role in the realization of unconventional FCIs. Why does the center-localized state appear in hyperbolic lattices? Geometry responses (such as the gravitational anomaly and electromagnetic response) have been revealed in the curved fractional quantum Hall states~\cite{Geometry1,Geometry_Yang1,Geometry2,Geometry3,Geometry_Yang2,Geometry4,Geometry_HHTu,Geometry5}. It is very interesting to explore geometry responses for these two types of hyperbolic FCIs. Additionally, non-Abelian FCIs have already been systematically studied in 2D Euclidean lattices~\cite{YFW2,ZLiu3,HeAL2,FCI_reviews,FCI_reviews1}, while non-Abelian FCIs in hyperbolic lattices are awaiting to be explored. FCIs have been verily observed in Moir{\'e} systems~\cite{Cai2023,Zeng2023,Park2023,XuFan2023,lu2023fractional}. The study of electronic properties and flat bands in possible hyperbolic Moir{\'e} systems will also be appealing. Experimentally, the realization of 2D hyperbolic lattices have been reported in both the cQED~\cite{Kollar2019,Boettcher} and the classical electric-circuit networks~\cite{TFBZhang,Lenggenhager2022,Chen2023,Zhang2022,Zhang2023,PeiQS}. Recently, the TFBs in OKG model have been realized in the electric circuits~\cite{TFBZhang}, however, it is very difficult to add interactions in the electric circuits. The cQED~\cite{Kollar2019,Boettcher} and cold-atomic systems can be candidates to realize hyperbolic FCIs in which both TFBs and interactions are possibly achieved.

{ {\it Acknowledgments}}---
This work was supported in part by the NSFC under Grants Nos. 12204404 (A.-L.H.), 12304557 (L.Q.) and 11874325 (Y.-F.W.), and the Natural Science Foundation of Jiangsu Higher Education Institutions of China, Grant No. 22KJB140019 (A.-L.H.).

\section*{APPENDIX}
\setcounter{section}{0}
\renewcommand \thesection{A\arabic{section}}

In main text, we investigated fractional Chern insulators (FCIs) in hyperbolic lattices which host constant negative curvature. When considering the hard-core bosons filling into the hyperbolic topological flat band (TFB) models, there are two types of $\nu=1/2$ FCIs, {\emph i.e.}, the conventional and the unconventional $\nu=1/2$ FCIs. Based on the exactly diagonalization (ED) results, two branches of edge excitations are found in these FCIs. Furthermore, trial wavefunctions (WFs) for the $\nu=1/2$ FCIs are constructed based on the single particle states, the generalized Pauli principle (GPP) and the Jack polynomials (Jacks). Based on the value of the WF overlap between the trial WFs and the ED results, geometric degree of freedom of both FCI states are revealed. However, the geometry makes various contributions to them.

\section{Topological flat bands in the octagon-kagome and the nonagon-kagome lattice models} \label{HTFBs}
\begin{figure}[!htb]
\includegraphics[scale=0.7]{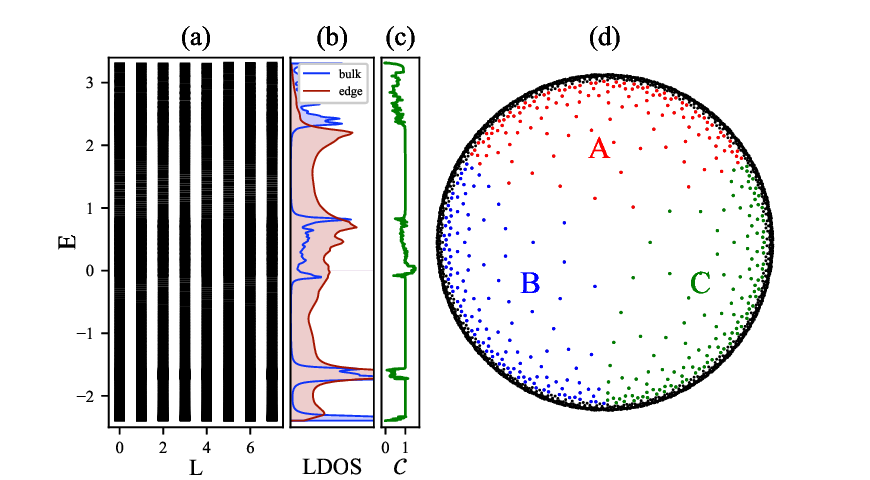}
\caption{(color online). (a) Single-particle energies of OKG lattice model are arranged in various angular momentum $L$ sectors with the hopping parameters of TFB, (b) the corresponding bulk and edge LDOS and (c) the real-space Chern number.  Here, the OKG model is with 3016 sites. (d) The partial bulk with 399 sites is divided into three regions respectively marked with ``A", ``B" and ``C" arranged in the counterclockwise order.}
\label{OKG1}
\end{figure}

TFBs are explored in the octagon-kagome (OKG) model. The OKG model consists of regular octagons and triangles, which hosts eightfold rotational symmetry. We still employ the hopping parameters of 2D Euclidean KG-TFB model which involves the nearest-neighbor (NN) and next-nearest-neighbor (NNN) hopping process and the corresponding hopping parameters are $t_1=-1.0$, $\phi=-0.22\pi$ and $t_2=0.19$~\cite{TFBKG1}. To obtain the energy eigenvalues, we diagonalize the single-particle Hamiltonian with open boundary conditions. These energy eigenvalues can be arrange in the angular momentum sectors marked with $L$ from ``0" to ``7" resulting from the eightfold rotational symmetry [in Fig.~\ref{OKG1} (a)]. We present the bulk and edge LDOS in Fig.~\ref{OKG1} (b) where the model has four bulk bands and the lowest band is relatively narrow implying the emergence of flat band. To characterize the topological properties, we calculate the real-space Chern number as functions of the fermi energy $E$ based on the Kitaev's formula. The bulk of HKG disk is cut into three distinct neighboring regions arranged in the counterclockwise order shown in Fig.~\ref{OKG1} (c) and the regions are marked with ``A", ``B" and ``C". Substantial plateau of quantized real-space Chern number appears when the fermi energy $E$ locates in the gap between the first and the second lowest bands [in Fig.~\ref{OKG1} (d)], which suggests that the lowest band belongs to a TFB.

\begin{figure}[!htb]
\includegraphics[scale=0.8]{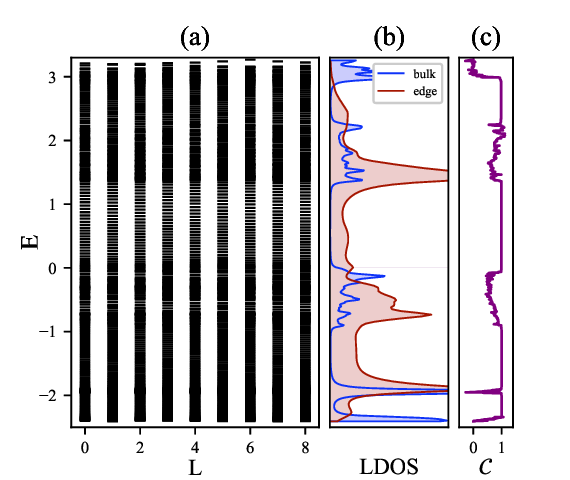}
\caption{(color online). (a) Single-particle energies of NKG lattice model are arranged in various angular momentum $L$ sectors with the hopping parameters of TFB, (b) the corresponding bulk and edge LDOS and (c) the real-space Chern number.  Here, the NKG model is with 2322 sites.}
\label{NKGEG0}
\end{figure}

TFB models can be extended to the nonagon-kagome (NKG) model as well.  The NKG model consists of regular nonagons and triangles with ninefold rotational symmetry. Here, we also adopt the hopping parameters of the 2D Euclidean KG-TFB model~\cite{TFBKG1}, $\emph i.e.$, $t_1=-1.0$, $t_2=0.19$ and $\phi=0.22$. Based on these hopping parameters, we obtain the single particle energy spectrum by diagonalizing the Hamiltonian with open boundary conditions. The energy eigenvalues can be arrange in each angular momentum sector marked with ``0" to ``8" [Fig.~\ref{NKGEG0} (a)] on account of the ninefold rotational symmetry. To demonstrate the existence of TFBs, we present the bulk and edge local density of state (LDOS) in Fig.~\ref{NKGEG0} (b) and the lowest energy band seems nearly flat with the divergent bulk LDOS. To characterize the topological feature of this near flat band, we calculate the real-space Chern number as functions of the fermi energy $E$ based on the Kitaev's formula  and substantial plateau of quantized real-space Chern number emerges when the fermi energy $E$ locates in the gap above the lowest energy band [Fig.~\ref{NKGEG0} (c)]. On the basis of these numerical results, the NKG model hosts a TFB.

\section{Single particle orbital and edge excitations in HKG-TFB model} \label{EEs1}
\begin{figure*}
\includegraphics[width=14cm]{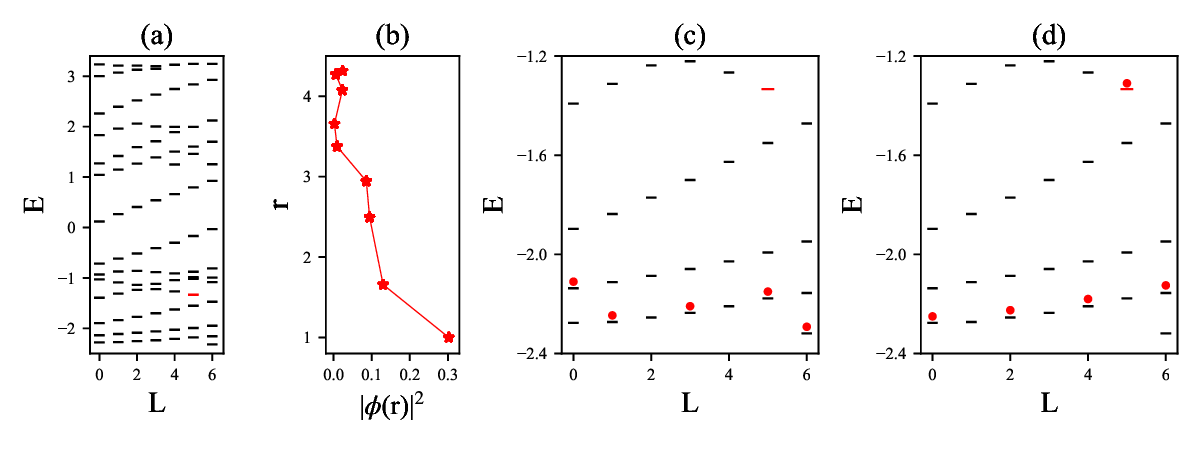}
\caption{(color online). (a) Single-particle spectra of the 98-site HKG-TFB model with trap potential $V_{\rm trap}=0.02$ versus the angular momentum quantum number $L$. Low-energy center-localized state is colored with red. (b). The density distribution of center-localized state along radial direction. Ground state configurations for (c) conventional and (d) unconventional $\nu=1/2$ FCIs in the HKG-TFB model. The low-energy center-localized orbital is colored with red and the red balls denote the bosons occupying in single-particle orbitals.}
\label{HKGEE}
\end{figure*}

The single-particle spectra can easily obtained by diagonalizing the Hamiltonian with a trap potential ($V_{\rm trap}=0.02$) with open boundary condition [displayed in Fig.~\ref{HKGEE} (a)]. On account of the HKG hosting sevenfold rotational symmetry, The single-particle energy spectra arranged in angular momentum sectors with angular momentum quantum number $L=0, 1, 2,..., 6$. Different from 2D Euclidean kagome lattice model, a state between the lowest and the second lowest bands [colored with red in Fig.~\ref{HKGEE} (b)] emerge whose WF distribution [shown in Fig.~\ref{HKGEE} (b)] is localized around the center of hyperbolic disk, and therefore it is dubbed as ``center-localized" state. The appearance of center-localized state attributes to geometry of HKG and it may be closely related to the non-Abelian Bloch state.

When hard-core bosons fills into this HKG-TFB model, we have obtained the EEs for $\nu=1/2$ FCIs.  Based on the GPP~\cite{GPP,GPP1,GPP2,GPP3}, these degeneracy sequences can be explained as multi branches of EEs. For example, the case of five hard-core bosons filling into HKG-TFB model, the GS root configuration of $\nu=1/2$ FCI is $|1010101010\rangle$ shown in Fig.~\ref{HKGEE} (c) and the corresponding angular momentum is $(6+1+3+5+0)$ $mod$ $7=1$, which is in accordance with the ED result in Fig.3 in the main text. Other root configurations of excited states can be obtained based on the GPP and the degeneracy sequence of these occupation configurations is ``1,1,2,3,5,7,...". On account of the existence of center-localized state, one particle can occupies in this orbital and eventually leads to another branch of EEs, analogue to the role of defect-core orbital in singular KG-TFB models. The root configuration of the initial many-body energy of this branch is $|\underline{1}01010101\rangle$, where $\underline{1}$ denotes one particle occupying the center-localized orbital [as shown in Fig.~\ref{HKGEE} (d)] and the corresponding angular momentum is $(5+0+2+4+6)$ $mod$ $7=3$. As a consequence, one branch of EEs with degeneracy sequence ``1,1,2,3,5,..." emerges from the initial angular momentum $L=3$. Combing with these two branches of EEs, the total degeneracy sequence becomes ``1,1,2,3,5,7,...+0,0,1,1,2,3,...=1,1,3,4,7,10,...".  One can easily obtain other degeneracy sequences of EEs with various number of particles based on the GPP.

\begin{figure*}
\includegraphics[width=14cm]{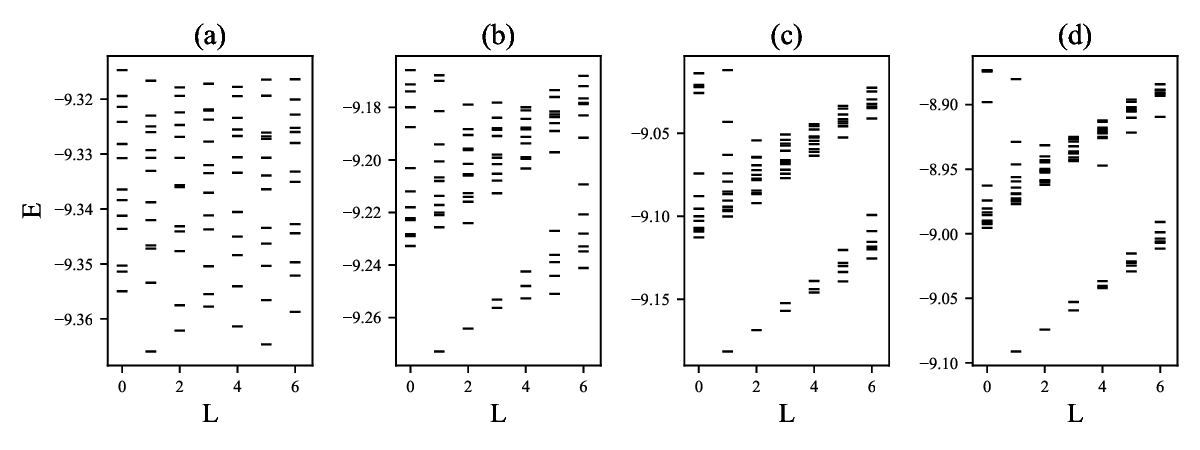}
\caption{(color online). EEs for FCIs in the 98-site HKG-TFB model with and without trap potentials $V_{trap}$. (a) $V_{trap}=0$. (b) $V_{trap}=0.005$. (c) $V_{trap}=0.01$. (d) $V_{trap}=0.015$. Here, we consider four hard-core bosons filling into the TFB model.}
\label{Clear_EES}
\end{figure*}

Additionally, trap potential plays key role to obtain the clear EEs for FCIs with open boundary conditions. Here, we take EEs of FCI in HKG-TFB model as an example and related numerical results are shown in Fig.~\ref{Clear_EES}. When without trap potential, there is no evidence to reveal the existence of EEs because of the absence of quasi-degeneracy sequences [in Fig.~\ref{Clear_EES} (a)]. When adding the trap potential, the quasi-degeneracy sequences emerge [Fig.~\ref{Clear_EES} (b)-(d)] and when the trap potential is $V_{trap}=0.015$, very clear EEs appear [Fig.~\ref{Clear_EES} (d)]. To obtain clear EEs for other numbers of hard-core bosons, we adopt $V_{trap}=0.02$ in the main text.

\section{Single particle orbital and edge excitations in OKG-TFB model} \label{EEs2}
\begin{figure*}
\includegraphics[width=14cm]{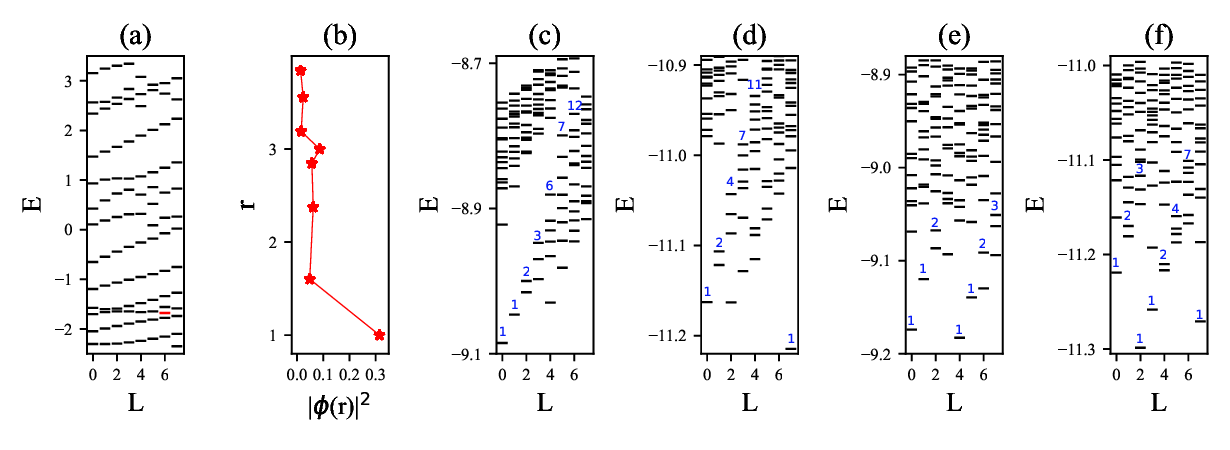}
\caption{(color online). (a) Single-particle spectra of the 104-site OKG-TFB model with trap potential $V_{\rm trap}=0.02$ versus the angular momentum quantum number $L$. Low-energy center-localized state is colored with red. (b). The density distribution of center-localized state along radial direction. EEs for the conventional $\nu=1/2$ FCI state with (c) four ($V=4.0$) and (d) five ($V=3.5$) hard-core bosons.  EEs for the unconventional $\nu=1/2$ FCI state with (e) four ($V=-0.7$) and (f) five ($V=0.0$) hard-core bosons. Numbers of EE quasidegeneracies in each sector are labelled upon the low-energy levels.}
\label{OKG2}
\end{figure*}

Based on the OKG-TFB model, we can explore the EEs for $\nu=1/2$ in hyperbolic disk with a harmonic trap potential $V_{\rm trap}=0.02$. The single-particle spectra can obtained by diagonalizing the Hamiltonian in every angular momentum sector [shown in Fig.~\ref{OKG2} (a)]. The angular momentum of GS is seven ($L=7$) and other low-energy orbitals arrange according to the angular momentum, {\emph i.e.}, the first excited state with $L=0$, the second excited state with $L=1$, ... [details in Fig.~\ref{OKG2} (a)]. Similar to single-particle state in the HKG-TFB model, there is a low-energy center-localized state [colored with red in Fig.~\ref{OKG2} (a)] which is localized around the center of the hyperbolic disk [as illustrated in Fig.~\ref{OKG2} (b)].

\begin{figure*}
\includegraphics[width=14cm]{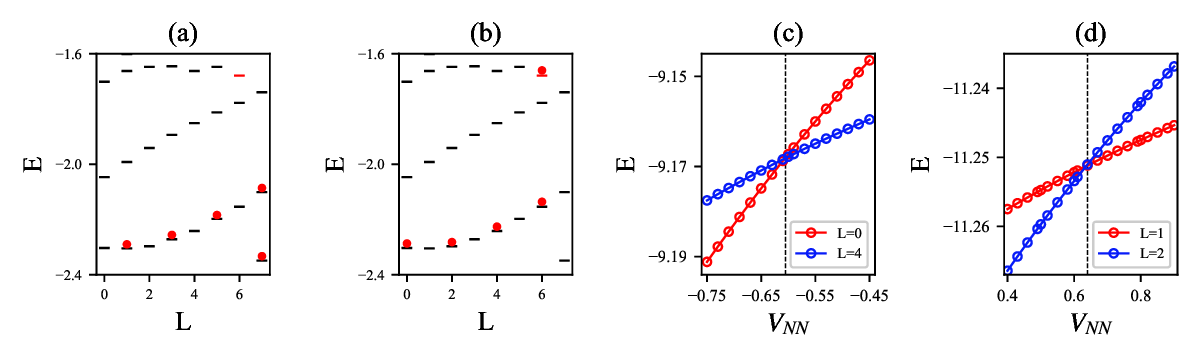}
\caption{(color online). GS configurations for conventional (a) and unconventional $\nu=1/2$ FCIs in the OKG-TFB model. The low-energy center-localized orbital is colored with red and the red balls denote the bosons occupying in single-particle orbitals. Various FCI states and energy crossing by adding the NN interaction $V_{NN}$ with four (c) and five (d) hard-core bosons filling the 104-site OKG-TFB model. The crossing points are near $V_{NN}\approx-0.81$ and $V_{NN}\approx-0.61$ for these two cases, respectively. $L$ denotes the angular momentum quantum number.}
\label{OKGEG0}
\end{figure*}

When hard-core bosons fill into this OKG-TFB model with the NN interaction, we can obtain EEs according to the ED method. Here, we introduce the NN repulsive interaction and consider four and five hard-core bosons filling in the OKG-TFB model. For the case of four hard-core bosons, the angular momentum of GS is zero [Fig.~\ref{OKG2} (c)] and its corresponding root configuration is probably ``$|10101010\rangle$" [displayed in Fig.~\ref{OKGEG0} (a)] whose total angular momentum is $(7+1+3+5)$ $mod$ 8=0. For the case of five hard-core bosons, the angular momentum of GS is seven [Fig.~\ref{OKG2} (c)] and the corresponding root configuration is probably ``$|1010101010\rangle$" [displayed in Fig.~\ref{OKGEG0} (a)] with total angular momentum $(7+1+3+5+7)$ $mod$ 8=7. For these root configurations, all particles occupy in the low-energy TFB orbitals, accordingly, these root configurations correspond to the conventional $\nu=1/2$ FCI. Subsequently, clear EE spectra with total degeneracy sequences ``1,1,2,3,6,7,12,..." for the case of four bosons [Fig.~\ref{OKG2} (c)] and ``1,1,2,4,7,11,..." for five bosons [Fig.~\ref{OKG2} (d)] emerge, which indicate that more than one branches of EE spectra mixed together for this $\nu=1/2$ FCI. According to the degeneracy sequences of $\nu=1/2$ in disk geometry~\cite{YFW4}, the new breeds of EEs with degeneracy sequences are ``1,1,3,..." (with four bosons, the initial angular momentum four) and  ``1,2,4,..." (with five bosons, the initial angular momentum two). Based on the GPP, one can identify that the root configuration for one branch of the mixed many-body state is $|\underline{1}01010101\rangle$, {\emph i.e.,} corresponding to the root configuration of unconventional $\nu=1/2$ FCI [as shown in Fig.~\ref{OKGEG0} (b)]. The total angular momenta for these initial energies are $(6+0+2+4)$ $mod$ 8 = 4 and $(6+0+2+4+6)$ $mod$ 8 = 2, in accordance with the ED results. However, for the unconventional $\nu=1/2$ FCI, the degeneracy sequence of EEs is still ``1,1,2,3,5,7,...". For this reason, there are additional EE spectra mixed together. Here, there may be two possible reasons for the appearance of the additional EE spectra. One is the EE spectra mixing with higher excited energies for the size effect of the finite lattice and the other is a branch of EEs with root configuration $|\underline{1}100001011\rangle$, in analogy to the $\nu=3/7$ FCI state. However, we are unable to provide a clear evidence for the emergence of the additional EE spectra.

When we tuning the strength of interaction, the GS energy angular momenta respectively change into $L=4$ and $L=2$ with four and five hard-core bosons [in Fig.~\ref{OKG2} (e) and (f)]. Subsequently, we find multi branches of EEs [in Fig.~\ref{OKG2} (e) and (f)] and the branch with GS energy hosts the degeneracy sequence ``1,1,2,3,5,..." which reveals the GS possibly belong to a $\nu=1/2$ unconventional FCI with the root configuration $|\underline{1}01010101\rangle$ where one particle occupies a center-localized orbital and the corresponding occupying configuration is shown in Fig.~\ref{OKGEG0} (b). Based on the GPP, one can obtain the angular momentum $(6+0+2+4)$ $mod$ 8 = 4 for four hard-core bosons and $(6+0+2+4+6)$ $mod$ 8 = 2 for five hard-core bosons which conforms to the ED results. Similar to the FCIs in the HKG TFB model, by tuning the NN interaction,  the unconventional  $\nu= 1/2$ FCI state can directly transition to the conventional $\nu= 1/2$ FCI state, without intermediate states as shown in Fig.~\ref{OKGEG0} (c) and (d).

\section{Single particle orbital and edge excitations in NKG-TFB model} \label{EEs3}
\begin{figure*}
\includegraphics[width=14cm]{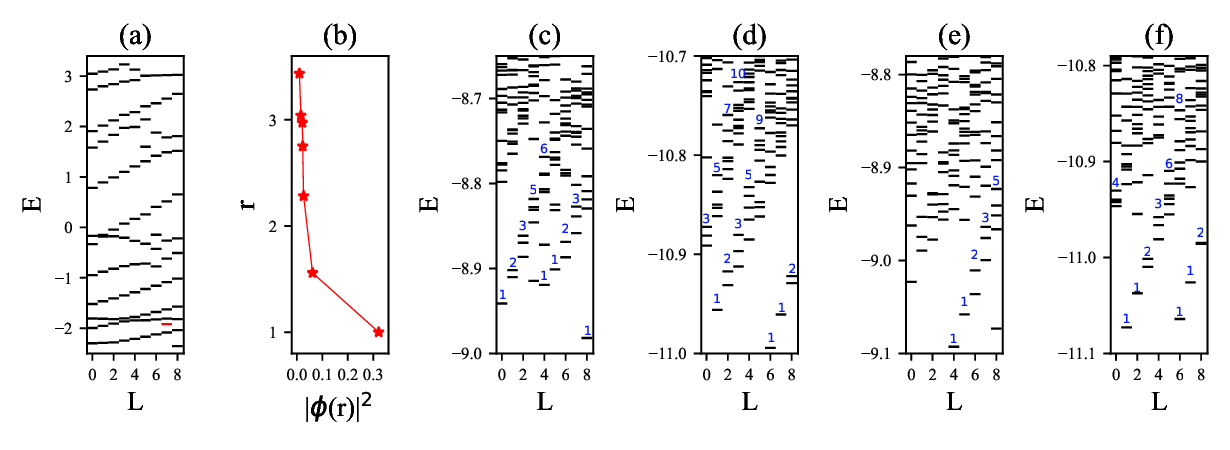}
\caption{(a) Single-particle spectra of the 108-site NKG-TFB model with trap potential $V_{\rm trap}=0.02$ versus the angular momentum quantum number $L$. Low-energy center-localized state is colored with red. (b). The density distribution of center-localized state along radial direction. EEs for the conventional $\nu=1/2$ FCI state with (c) four ($V_{\rm NN}=V_{\rm NNN}=1.0$) and (d) five ($V_{\rm NN}=V_{\rm NNN}=3.0$) hard-core bosons.  EEs for the unconventional $\nu=1/2$ FCI state with (e) four ($V_{\rm NN}=V_{\rm NNN}=0.0$) and (f) five ($V_{\rm NN}=V_{\rm NNN}=1.0$) hard-core bosons. Numbers of EE quasidegeneracies in each sector are labelled upon the low-energy levels.}
\label{NKGEG1}
\end{figure*}

\begin{figure*}
\includegraphics[width=14cm]{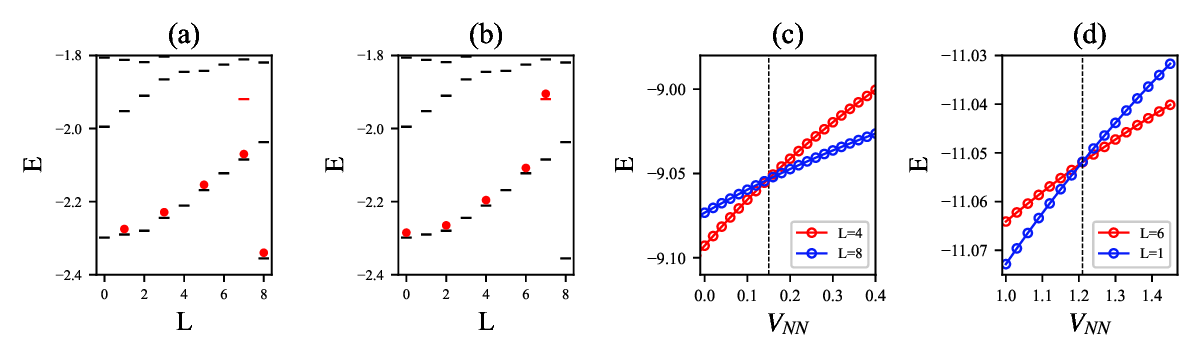}
\caption{GS configurations for conventional (a) and unconventional $\nu=1/2$ FCIs in the NKG-TFB model. The low-energy center-localized orbital is colored with red and the red balls denote the bosons occupying in single-particle orbitals. Various FCI states and energy crossing by adding the NN interaction $V_{NN}$ with four (c) and five (d) hard-core bosons filling the 108-site NKG-TFB model. The crossing points are near $V_{NN}\approx0.15$ and $V_{NN}\approx1.21$ for these two cases, respectively. $L$ denotes the angular momentum quantum number.}
\label{NKGEG2}
\end{figure*}

$\nu=1/2$ FCI states can be investigated based on this TFB model. Here, we consider a 108-site NKG disk with  a harmonic trap (trap potential $V_{trap}$= 0.02). By diagonalizing the Hamiltonian in each angular momentum sector, the single-particle spectra can be obtained as displayed in Fig.~\ref{NKGEG1} (a). The
angular momentum of GS is eight ($L = 8$) and other low-energy orbitals arrange according to the angular momentum, in analogy to the cases of HKG and OKG TFB models.  Simultaneously, low energy center-localized orbital emerges in the lowest bulk band [Fig.~\ref{NKGEG1} (a) colored with red] and the corresponding state is localized around the center of the NKG disk [in Fig.~\ref{NKGEG1} (b)].

Considering hard-core bosons filling into the NKG model,  $\nu=1/2$ FCI state can be obtained based on the real-space ED method. Here, we introduce the NN and the NNN interactions for NKG-TFB model with four and five hard-core bosons and we set $V_{\rm NN}=V_{\rm NNN}=V$. Clear EE spectra are respectively shown in Fig.~\ref{NKGEG1} (c) and (d) where multi branches of EEs appears and some branches of EE spectra mix together as well. For the case of four hard-core bosons, the angular momentum of GS is eight based on the ED result [Fig.~\ref{NKGEG1} (c)] and the corresponding root configuration is probably ``$|10101010\rangle$" with $V=1.0$ [occupation configuration in Fig.~\ref{NKGEG2} (a)] whose total angular momentum is $(8+1+3+5)$ $mod$ 9 =8. For the case of five hard-core bosons, the angular momentum of GS is six [Fig.~\ref{NKGEG1} (c)] and the corresponding root configuration is probably ``$|1010101010\rangle$" with $V=1.0$ [occupation configuration in Fig.~\ref{NKGEG2} (a)] whose total angular momentum is $(8+1+3+5+7)$ $mod$ 9 =6. This $\nu=1/2$ FCI state belong to the conventional $\nu=1/2$ FCI state because all particles occupy into the low-energy TFB orbitals and no particle occupy into the center-localized orbital. We can find two branches of EE spectra with degeneracy sequences ``1,1,2,3,5,6..." and ``1,1,2,3,..." for the case of four bosons, and two branches of EE spectra with degeneracy sequences ``1,1,2,3,5,7,10" and ``1,2,3,5,9,..." for the case of five bosons. Based on the results of EE spectra for the HKG-TFB and OKG-TFB models, one can verify that one branch of EE spectra belong to the unconventional $\nu=1/2$ FCI which one boson occupies the center-localized orbital [the occupation configuration in Fig.~\ref{NKGEG2} (b)]. For the FCIs in the NKG tFB model, by tuning the NN interaction,  the unconventional  $\nu= 1/2$ FCI state can directly transition to the conventional $\nu= 1/2$ FCI state, without intermediate states as shown in Fig.~\ref{NKGEG2} (c) and (d).

\section{The method of constructing trial WFs for FCIs based on the GPP} \label{WF0}
Finding the optimal trial WFs for FCI states becomes an outstanding problem. The first trial WFs for FCIs were constructed based on the mapping relationship between the orbitals in LLs and the  maximally localized Wannier functions defined in CIs on cylinder geometry~\cite{Qi}. Subsequently, an improved prescription has been adopted to construct variational WFs for FCIs on torus geometry by adopting the gauge-fixed (non-maximally) localized Wannier~\cite{YLWu,YLWu0}. By mapping single-particle orbitals of TFBs into the ones in the LLs, trial WFs for both Abelian and non-Abelian FCIs in disk geometry have been reported~\cite{HeAL1,HeAL2}. Simultaneously, developing this direct and effective approach to obtain trial WFs is inspired by the polynomial structure of WFs for FQH states, which connotes the GPP.

Some WFs for FQH states host analytical form, for example, the Laughlin WFs with filling factor $\nu=1/m$ in disk geometry with this expression $\Psi_{\rm {FQH}}^{\nu=1/m}(\{z_i\})=\prod_{i<j} (z_i-z_j)^m {\rm exp}(-\sum_i|z_i|^2/4)$~\cite{Laughlin1983}, where $z_i=x_i+iy_i$ is a complex coordinate. Here, we take the $\nu=1/2$ Laughlin WF with 3 bosons as an example, {\emph {i.e.},} $\Psi_{\rm{FQH}}^{\nu=1/2}(z_1,z_2,z_3)= (z_1-z_2)^2 (z_1-z_3)^2 (z_2-z_3)^2$. For simplicity, we drop the non-universal Gaussian factor ${\rm exp}(-\sum_i|z_i|^2/4)$.
One can expand $\Psi_{1/2}(z_1,z_2,z_3)$ as,
\begin{eqnarray}
&&\Psi_{1/2}(z_1,z_2,z_3)=(+1)({z_1}^{4} { z_2}^{2} {z_3}^{0}+{z_1}^{4}{z_2}^{0}{z_3}^{2}+{ z_1}^{2} { z_2}^{4}{ z_3}^{0} \nonumber\\&& +{z_1}^{2}{z_2}^{0}{z_3}^{4} + {z_1}^{0}
{z_2}^{4}{ z_3}^{2}+{ z_1}^{0}{ z_2}^{2}{ z_3}^{4}) + (-2)({z_1}^{4}{z_2}^{1}{ z_3}^{1} + \nonumber\\&& {z_1}^{1} { z_2}^{4}{z_3}^{1}+
{ z_1}^{1}{z_2}^{1}{z_3}^{4} )+(-2)({z_1}^{3}{z_2}^{3} {z_3}^{0}+{z_1}^{3}{ z_2}^{0}{ z_3}^{3}  + \nonumber\\&& { z_1}^{0}{z_2}^{3}{z_3}^{3})
+(+2) ({z_1}^{3}{ z_2}^{2}{ z_3}^{1}+{ z_1}^{3}{ z_2}^{1}{ z_3}^{2} +{z_1}^{1}{z_2}^{2} {z_3}^{3}+ \nonumber\\&& { z_1}^{1}{ z_2}^{3}{z_3}^{2} +{ z_1}^{2}{ z_2}^{1}{z_3}^{3}
+{z_1}^{2} { z_2}^{3}{z_3}^{1})+ (-6){(z_1}^{2}{ z_2}^{2}{z_3}^{2})
\label{Expand_polynomial}
\end{eqnarray}
The above expansion can be simplified as
\begin{eqnarray}
&&\Psi_{\rm{FQH}}^{\nu=1/2}(z_1,z_2,z_3)=(+1)\Phi_{[4,2,0]}+(-2)\Phi_{[4,1,1]}+ \nonumber\\&& (-2)\Phi_{[3,3,0]}+(+2)\Phi_{[3,2,1]}+(-6)\Phi_{[2,2,2]},
\label{Expand_polynomial1}
\end{eqnarray}
where $\Phi_{[4,2,0]}$ denotes ${z_1}^{4} { z_2}^{2} {z_3}^{0}+{z_1}^{4}{z_2}^{0}{z_3}^{2}+{ z_1}^{2} { z_2}^{4}{ z_3}^{0} +{z_1}^{2}{z_2}^{0}{z_3}^{4} + {z_1}^{0}{z_2}^{4}{ z_3}^{2}+{ z_1}^{0}{ z_2}^{2}{ z_3}^{4}$. The WF for the lowest LL is $\phi_m (z) \propto z^m$, therefore, $\Phi_{[4,2,0]}$ and others are the many-particle wfs for free bosons, ${\emph {i.e.}}$, $\Phi_{[4,2,0]}\propto \phi_4 (z_1)\phi_2 (z_2)\phi_0 (z_3) +  \phi_4 (z_1)\phi_0 (z_2)\phi_2 (z_3) + \phi_2 (z_1)\phi_4 (z_2)\phi_0 (z_3) + \phi_2 (z_1)\phi_0 (z_2)\phi_4 (z_3) + \phi_0 (z_1)\phi_4 (z_2)\phi_2 (z_3) + \phi_0 (z_1)\phi_2 (z_2)\phi_4 (z_3) $. $\Phi_{[4,2,0]}$ is defined as the root configuration. Consequently, one can find the Laughlin WF is decomposed as,
\begin{eqnarray}
\Psi_{\rm{FQH}}^{\nu=1/m}({z_i})=\sum_l J_{\lambda_l} \Phi_{\lambda_l} ^{\rm LLL}.
\label{Expand_polynomial2}
\end{eqnarray}
 Here $\Phi^{\rm LLL}_{\lambda_l}$ is the antisymmetric Slater determinant (for fermions) or symmetric polynomial (for bosons) composed by the single-particle states of the lowest LL, the expansion coefficients $J_{\lambda_l}$ can be obtained  by recurrence relation  and  the $l-$th basis configurations $\lambda_l$ comes from squeezing the root configuration~\cite{Bernevig1,Bernevig2,Bernevig3}.  Other WFs can be decomposed, and the basis configurations $\lambda_l$ and corresponding expansion coefficients $J_{\lambda_l}$ can be obtained with the aid of the Jack polynomials (Jacks)~\cite{Bernevig1,Bernevig2,Bernevig3}.

The distributions of low-energy single-particle states of KG-TFB model in disk geometry are similar to the ones in the lowest LL .
Accordingly, the mapping relationship between single-particle orbitals of KG-TFB model and the ones for LLs can be established. For example, the single-particle state with $m=0$ and the first low-energy state in the KG-TFB, the single-particle state with $m=1$ and the second low-energy state in the KG-TFB, etc. For $\nu=1/2$ FCI state, one can substitute single-particle states of KG-TFB model  for the WFs with angular momentum quantum numbers $m$ in the lowest LL and put them into Eqs.~(\ref{Expand_polynomial1}) and ~(\ref{Expand_polynomial2}). The trial WF for $\nu=1/2$ FCI can be directly obtained~\cite{HeAL3}, ${\emph i.e.},$
\begin{eqnarray}
\Psi_{\rm{FCI}}^{\nu=1/2}=\sum_l J_{\lambda_l} \Phi_{\lambda_l} ^{\rm TFB}.
\label{Expand_polynomial3}
\end{eqnarray}
Here, $\Phi_{\lambda_l} ^{\rm TFB}$ can be calculated based on the single-particle states of KG-TFB. Especially note that the trial WFs of Eq.~(\ref{Expand_polynomial3}) belongs to a soft-core bosonic WF, instead of hard-core case. For hard-core bosons, no more than one particle occupies in the same lattice site. Consequently, the soft-core WF should be projected into the Hilbert space hard-core state~\cite{HeAL3}, ${\emph i.e.},$
\begin{eqnarray}
\Psi_{\rm{FCI, HC}}^{\nu=1/2}= {\cal N} {\hat {\cal P}} \Psi_{\rm{FCI}}^{\nu=1/2}={\cal N} \prod_{i,j} (1-\delta_{i,j}) \Psi_{\rm{FCI}}^{\nu=1/2},
\label{Expand_polynomial4}
\end{eqnarray}
where $\Psi_{\rm{FCI, HC}}^{\nu=1/2}$ is the trial WF for $\nu=1/2$ FCI filling with hard-core bosons and ${\hat {\cal P}}=\prod_{i,j} (1-\delta_{i,j})$ is the projection operator depending on the Kronecker delta function $\delta_{i,j}$ with $i$ and $j$ marking the site positions. For fermionic FCIs or bosonic FCIs filling with soft-core bosons, this projection process does not require.
To estimate the rationality of this trial WF, a WF overlap between the trial one and the ED result. The value of WF overlap is higher than 0.98 even when five hard-core bosons occupy the KG-TFB model and the corresponding Hilbert space dimension is as large as $8\times10^5$. High value of WF overlap manifests the feasibility of this method. Trial WFs for non-Abelian FCIs in disk geometry have been successfully constructed based on this method. We also investigated $\nu=1/2$ FCIs in singular TFB models and reveal the interplay of geometry and FCI states.

\begin{figure}[!htb]
\includegraphics[scale=0.7]{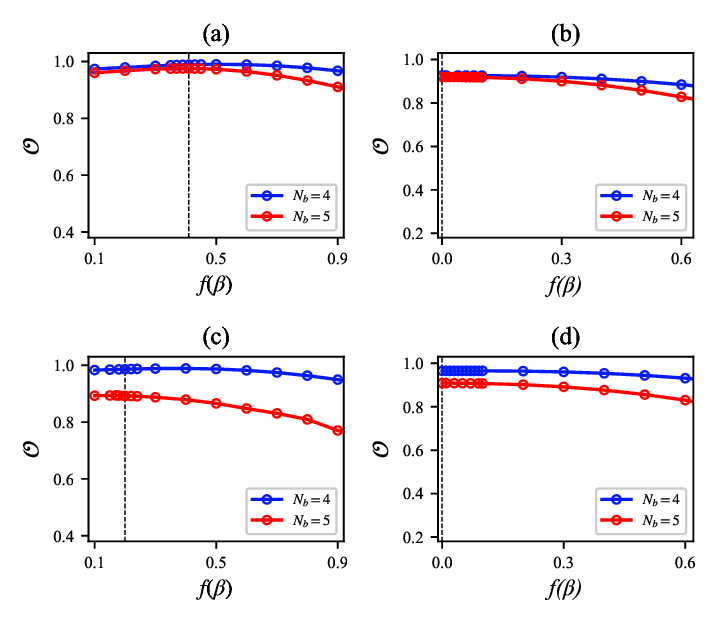}
\caption{WF Overlap between the trial WFs and the ED results with variable geometric function $f(\beta)$ for $\nu=1/2$ FCIs filling with $N_b=4$ and $N_b=5$ hard-core bosons in (a), (b) OKG, and (c), (d) NKG TFB models. Here, (a) and (c) denote the results of the conventional $\nu=1/2$ FCI and, (d) and (d) show the overlap results for the unconventional $\nu=1/2$ FCI. The vertical lines denote the suitable $f(\beta)$ corresponding to the maximum WF overlap values.}
\label{OL}
\end{figure}

Different from the WF for $\nu=1/2$ FQH state in disk geometry, the conical version of $\nu=1/2$ FQH state is,
\begin{equation}\label{Bose_Laughlin}
\Psi^{\nu=1/2}_{\rm{FQH}}(\{z_i\},\beta)=\displaystyle\prod_{i<j}(z^{\beta}_i-z^{\beta}_j)^{2}{\rm{exp}}({-\displaystyle\sum_i|z_i|^2/4}),
\end{equation}
where $\beta$ is geometric factor which marks the remaining part of a disk after cutting. The single-particle WF for the lowest LL is $\phi_m(z,\beta)\propto (z^\beta)^m$. One can find $\Psi_{\rm{FQH}}(\{z_i\},\beta)$ can be decomposed as,
\begin{eqnarray}
\Psi_{\rm{FQH}}^{\nu=1/2}({z_i},\beta)=\sum_l J_{\lambda_l}(\beta) \Phi_{\lambda_l} ^{\rm LLL} (\beta),
\label{Expand_polynomial2x}
\end{eqnarray}
analogue to FQH in disk geometry [Eq.~(\ref{Expand_polynomial2})], but $J_{\lambda_l}(\beta)$ is related to the geometric factor.
We can construct trial WFs for the KG-singular lattice with $n-$fold rotational symmetry by substituting the single-particle states of singular KG-TFB model for
the WFs of the lowest LL in conical surface~\cite{HeAL3}, {\emph i.e.},
\begin{eqnarray}
\Psi_{\rm{FCI}}^{\nu=1/2}(\beta)=\sum_l J_{\lambda_l} (\beta) \Phi_{\lambda_l} ^{\rm TFB} (\beta).
\label{Expand_polynomial3x}
\end{eqnarray}
Here, the geometric factor for the singular kagome lattice is $\beta_n=6/n$. After the projection [Eq.~(\ref{Expand_polynomial4})], the trial WF for singular $\nu=1/2$ FCI is constructed. There are two types of $\nu=1/2$ FCI states in KG-singular TFB models: i) one is the conventional $1/2$ FCI state where particles occupy in the low-energy orbitals and ii) the other is the unconventional $1/2$ FCI state where one particle occupies the defect-core orbital~\cite{HeAL3,HeAL4}. We found the maximum values of the WF overlap are 0.979 and 0.937 for these two types of FCIs with five bosons~\cite{HeAL3}. Interestingly, when the the WF overlap value become maximum, the corresponding geometric factor is $\beta=\beta_8=6/8$, which reveals the geometry plays the same role in these two types of $\nu=1/2$ FCIs~\cite{HeAL3}.

\section{WF values for $\nu=1/2$ FCIs in hyperbolic kagome TFB models} \label{WF1}
We have constructed the WFs for $\nu=1/2$ FCIs in hyperbolic TFB models with a general expression, {\emph i.e.,}
\begin{eqnarray}
\Psi_{\rm{HFCI}}^{\nu=1/2}=\sum_l J_{\lambda_l} ( f(\beta)) \Phi_{\lambda_l} ^{\rm TFB} (\beta).
\label{Expand_polynomialHT}
\end{eqnarray}
Here, ${\rm HFCI}$ denotes the FCIs in hyperbolic TFB models, $f(\beta)$ is the function of $\beta$ and $\beta=6/n$ for the hyperbolic model with $n-$fold rotational symmetry. Therefore, based on the single-particle states of the hyperbolic TFB models, the trial WFs can be obtained. We have taken the trial WFs for $\nu=1/2$ FCIs in the HKG TFB models as an example in the main text. Here, we present some results for the remaining hyperbolic TFB models mentioned in the main text (details shown in Fig.~\ref{OL}). For $\nu=1/2$ FCIs in the 72-site OKG TFB model, the maximum WF overlap values appear at $f(\beta)\approx 0.41$ for the conventional $\nu=1/2$ FCI state and $f(\beta)\approx 0.0$ for the unconventional $\nu=1/2$ FCI state based on the ED results. For this conventional $\nu=1/2$ FCI state, we add the NN interactions $V_{NN}=2.0$ for the case of four hard-core bosons and $V_{NN}=3.6$ for the case of five hard-core bosons. For this unconventional $\nu=1/2$ FCI state, no NN interaction is added. For $\nu=1/2$ FCIs in the 90-site NKG TFB model, the maximum WF overlap values appear at $f(\beta)\approx 0.20$ for the conventional $\nu=1/2$ FCI state and $f(\beta)$ very close to zero for the unconventional $\nu=1/2$ FCI state.  For this conventional $\nu=1/2$ FCI state, we add the NN and NNN interactions $V_{NN}=V_{NNN}=0.4$ for the case of four hard-core bosons and $V_{NN}=V_{NNN}=2.7$ for the case of five hard-core bosons. For the unconventional $\nu=1/2$ FCI state, no NN and NNN interactions are added ($V_{NN}=V_{NNN}=0.0$) for four and five bosons filling. Additionally, the WF overlaps for the conventional $\nu=1/2$ FCI state with five bosons filling are not very high (about 0.90) [shown in Fig.~\ref{OL} (c)] because there are only about ten orbitals in the lowest energy bands and lowest-energy orbitals are very close to the second lowest energy bands [details in Fig.~\ref{NKGEG1} (a)].

\nocite{*}
\bibliography{HyperFCI}

\begin{thebibliography}{94}%
\makeatletter
\providecommand \@ifxundefined [1]{%
 \@ifx{#1\undefined}
}%
\providecommand \@ifnum [1]{%
 \ifnum #1\expandafter \@firstoftwo
 \else \expandafter \@secondoftwo
 \fi
}%
\providecommand \@ifx [1]{%
 \ifx #1\expandafter \@firstoftwo
 \else \expandafter \@secondoftwo
 \fi
}%
\providecommand \natexlab [1]{#1}%
\providecommand \enquote  [1]{``#1''}%
\providecommand \bibnamefont  [1]{#1}%
\providecommand \bibfnamefont [1]{#1}%
\providecommand \citenamefont [1]{#1}%
\providecommand \href@noop [0]{\@secondoftwo}%
\providecommand \href [0]{\begingroup \@sanitize@url \@href}%
\providecommand \@href[1]{\@@startlink{#1}\@@href}%
\providecommand \@@href[1]{\endgroup#1\@@endlink}%
\providecommand \@sanitize@url [0]{\catcode `\\12\catcode `\$12\catcode `\&12\catcode `\#12\catcode `\^12\catcode `\_12\catcode `\%12\relax}%
\providecommand \@@startlink[1]{}%
\providecommand \@@endlink[0]{}%
\providecommand \url  [0]{\begingroup\@sanitize@url \@url }%
\providecommand \@url [1]{\endgroup\@href {#1}{\urlprefix }}%
\providecommand \urlprefix  [0]{URL }%
\providecommand \Eprint [0]{\href }%
\providecommand \doibase [0]{https://doi.org/}%
\providecommand \selectlanguage [0]{\@gobble}%
\providecommand \bibinfo  [0]{\@secondoftwo}%
\providecommand \bibfield  [0]{\@secondoftwo}%
\providecommand \translation [1]{[#1]}%
\providecommand \BibitemOpen [0]{}%
\providecommand \bibitemStop [0]{}%
\providecommand \bibitemNoStop [0]{.\EOS\space}%
\providecommand \EOS [0]{\spacefactor3000\relax}%
\providecommand \BibitemShut  [1]{\csname bibitem#1\endcsname}%
\let\auto@bib@innerbib\@empty
\bibitem [{\citenamefont {Neupert}\ \emph {et~al.}(2011)\citenamefont {Neupert}, \citenamefont {Santos}, \citenamefont {Chamon},\ and\ \citenamefont {Mudry}}]{Neupert}%
  \BibitemOpen
  \bibfield  {author} {\bibinfo {author} {\bibfnamefont {T.}~\bibnamefont {Neupert}}, \bibinfo {author} {\bibfnamefont {L.}~\bibnamefont {Santos}}, \bibinfo {author} {\bibfnamefont {C.}~\bibnamefont {Chamon}},\ and\ \bibinfo {author} {\bibfnamefont {C.}~\bibnamefont {Mudry}},\ }\bibfield  {title} {\bibinfo {title} {Fractional quantum hall states at zero magnetic field},\ }\href {https://doi.org/10.1103/PhysRevLett.106.236804} {\bibfield  {journal} {\bibinfo  {journal} {Phys. Rev. Lett.}\ }\textbf {\bibinfo {volume} {106}},\ \bibinfo {pages} {236804} (\bibinfo {year} {2011})}\BibitemShut {NoStop}%
\bibitem [{\citenamefont {Tang}\ \emph {et~al.}(2011)\citenamefont {Tang}, \citenamefont {Mei},\ and\ \citenamefont {Wen}}]{Tang}%
  \BibitemOpen
  \bibfield  {author} {\bibinfo {author} {\bibfnamefont {E.}~\bibnamefont {Tang}}, \bibinfo {author} {\bibfnamefont {J.-W.}\ \bibnamefont {Mei}},\ and\ \bibinfo {author} {\bibfnamefont {X.-G.}\ \bibnamefont {Wen}},\ }\bibfield  {title} {\bibinfo {title} {High-temperature fractional quantum hall states},\ }\href {https://doi.org/10.1103/PhysRevLett.106.236802} {\bibfield  {journal} {\bibinfo  {journal} {Phys. Rev. Lett.}\ }\textbf {\bibinfo {volume} {106}},\ \bibinfo {pages} {236802} (\bibinfo {year} {2011})}\BibitemShut {NoStop}%
\bibitem [{\citenamefont {Sheng}\ \emph {et~al.}(2011)\citenamefont {Sheng}, \citenamefont {Gu}, \citenamefont {Sun},\ and\ \citenamefont {Sheng}}]{Sheng1}%
  \BibitemOpen
  \bibfield  {author} {\bibinfo {author} {\bibfnamefont {D.~N.}\ \bibnamefont {Sheng}}, \bibinfo {author} {\bibfnamefont {Z.-C.}\ \bibnamefont {Gu}}, \bibinfo {author} {\bibfnamefont {K.}~\bibnamefont {Sun}},\ and\ \bibinfo {author} {\bibfnamefont {L.}~\bibnamefont {Sheng}},\ }\bibfield  {title} {\bibinfo {title} {Fractional quantum hall effect in the absence of landau levels},\ }\href {https://doi.org/10.1038/ncomms1380} {\bibfield  {journal} {\bibinfo  {journal} {Nature Communications}\ }\textbf {\bibinfo {volume} {2}},\ \bibinfo {pages} {389 EP } (\bibinfo {year} {2011})},\ \bibinfo {note} {article}\BibitemShut {NoStop}%
\bibitem [{\citenamefont {Wang}\ \emph {et~al.}(2011)\citenamefont {Wang}, \citenamefont {Gu}, \citenamefont {Gong},\ and\ \citenamefont {Sheng}}]{YFW1}%
  \BibitemOpen
  \bibfield  {author} {\bibinfo {author} {\bibfnamefont {Y.-F.}\ \bibnamefont {Wang}}, \bibinfo {author} {\bibfnamefont {Z.-C.}\ \bibnamefont {Gu}}, \bibinfo {author} {\bibfnamefont {C.-D.}\ \bibnamefont {Gong}},\ and\ \bibinfo {author} {\bibfnamefont {D.~N.}\ \bibnamefont {Sheng}},\ }\bibfield  {title} {\bibinfo {title} {Fractional quantum hall effect of hard-core bosons in topological flat bands},\ }\href {https://doi.org/10.1103/PhysRevLett.107.146803} {\bibfield  {journal} {\bibinfo  {journal} {Phys. Rev. Lett.}\ }\textbf {\bibinfo {volume} {107}},\ \bibinfo {pages} {146803} (\bibinfo {year} {2011})}\BibitemShut {NoStop}%
\bibitem [{\citenamefont {Regnault}\ and\ \citenamefont {Bernevig}(2011)}]{Regnault1}%
  \BibitemOpen
  \bibfield  {author} {\bibinfo {author} {\bibfnamefont {N.}~\bibnamefont {Regnault}}\ and\ \bibinfo {author} {\bibfnamefont {B.~A.}\ \bibnamefont {Bernevig}},\ }\bibfield  {title} {\bibinfo {title} {Fractional chern insulator},\ }\href {https://doi.org/10.1103/PhysRevX.1.021014} {\bibfield  {journal} {\bibinfo  {journal} {Phys. Rev. X}\ }\textbf {\bibinfo {volume} {1}},\ \bibinfo {pages} {021014} (\bibinfo {year} {2011})}\BibitemShut {NoStop}%
\bibitem [{\citenamefont {Qi}(2011)}]{Qi}%
  \BibitemOpen
  \bibfield  {author} {\bibinfo {author} {\bibfnamefont {X.-L.}\ \bibnamefont {Qi}},\ }\bibfield  {title} {\bibinfo {title} {Generic wave-function description of fractional quantum anomalous hall states and fractional topological insulators},\ }\href {https://doi.org/10.1103/PhysRevLett.107.126803} {\bibfield  {journal} {\bibinfo  {journal} {Phys. Rev. Lett.}\ }\textbf {\bibinfo {volume} {107}},\ \bibinfo {pages} {126803} (\bibinfo {year} {2011})}\BibitemShut {NoStop}%
\bibitem [{\citenamefont {Bernevig}\ and\ \citenamefont {Regnault}(2012)}]{GPP}%
  \BibitemOpen
  \bibfield  {author} {\bibinfo {author} {\bibfnamefont {B.~A.}\ \bibnamefont {Bernevig}}\ and\ \bibinfo {author} {\bibfnamefont {N.}~\bibnamefont {Regnault}},\ }\bibfield  {title} {\bibinfo {title} {Emergent many-body translational symmetries of abelian and non-abelian fractionally filled topological insulators},\ }\href {https://doi.org/10.1103/PhysRevB.85.075128} {\bibfield  {journal} {\bibinfo  {journal} {Phys. Rev. B}\ }\textbf {\bibinfo {volume} {85}},\ \bibinfo {pages} {075128} (\bibinfo {year} {2012})}\BibitemShut {NoStop}%
\bibitem [{\citenamefont {Wang}\ \emph {et~al.}(2012{\natexlab{a}})\citenamefont {Wang}, \citenamefont {Yao}, \citenamefont {Gu}, \citenamefont {Gong},\ and\ \citenamefont {Sheng}}]{YFW2}%
  \BibitemOpen
  \bibfield  {author} {\bibinfo {author} {\bibfnamefont {Y.-F.}\ \bibnamefont {Wang}}, \bibinfo {author} {\bibfnamefont {H.}~\bibnamefont {Yao}}, \bibinfo {author} {\bibfnamefont {Z.-C.}\ \bibnamefont {Gu}}, \bibinfo {author} {\bibfnamefont {C.-D.}\ \bibnamefont {Gong}},\ and\ \bibinfo {author} {\bibfnamefont {D.~N.}\ \bibnamefont {Sheng}},\ }\bibfield  {title} {\bibinfo {title} {Non-abelian quantum hall effect in topological flat bands},\ }\href {https://doi.org/10.1103/PhysRevLett.108.126805} {\bibfield  {journal} {\bibinfo  {journal} {Phys. Rev. Lett.}\ }\textbf {\bibinfo {volume} {108}},\ \bibinfo {pages} {126805} (\bibinfo {year} {2012}{\natexlab{a}})}\BibitemShut {NoStop}%
\bibitem [{\citenamefont {Parameswaran}\ \emph {et~al.}(2012)\citenamefont {Parameswaran}, \citenamefont {Roy},\ and\ \citenamefont {Sondhi}}]{Parameswaran}%
  \BibitemOpen
  \bibfield  {author} {\bibinfo {author} {\bibfnamefont {S.~A.}\ \bibnamefont {Parameswaran}}, \bibinfo {author} {\bibfnamefont {R.}~\bibnamefont {Roy}},\ and\ \bibinfo {author} {\bibfnamefont {S.~L.}\ \bibnamefont {Sondhi}},\ }\bibfield  {title} {\bibinfo {title} {Fractional chern insulators and the ${W}_{\ensuremath{\infty}}$ algebra},\ }\href {https://doi.org/10.1103/PhysRevB.85.241308} {\bibfield  {journal} {\bibinfo  {journal} {Phys. Rev. B}\ }\textbf {\bibinfo {volume} {85}},\ \bibinfo {pages} {241308} (\bibinfo {year} {2012})}\BibitemShut {NoStop}%
\bibitem [{\citenamefont {Wu}\ \emph {et~al.}(2012)\citenamefont {Wu}, \citenamefont {Regnault},\ and\ \citenamefont {Bernevig}}]{YLWu}%
  \BibitemOpen
  \bibfield  {author} {\bibinfo {author} {\bibfnamefont {Y.-L.}\ \bibnamefont {Wu}}, \bibinfo {author} {\bibfnamefont {N.}~\bibnamefont {Regnault}},\ and\ \bibinfo {author} {\bibfnamefont {B.~A.}\ \bibnamefont {Bernevig}},\ }\bibfield  {title} {\bibinfo {title} {Gauge-fixed wannier wave functions for fractional topological insulators},\ }\href {https://doi.org/10.1103/PhysRevB.86.085129} {\bibfield  {journal} {\bibinfo  {journal} {Phys. Rev. B}\ }\textbf {\bibinfo {volume} {86}},\ \bibinfo {pages} {085129} (\bibinfo {year} {2012})}\BibitemShut {NoStop}%
\bibitem [{\citenamefont {Wang}\ \emph {et~al.}(2012{\natexlab{b}})\citenamefont {Wang}, \citenamefont {Yao}, \citenamefont {Gong},\ and\ \citenamefont {Sheng}}]{WYF3}%
  \BibitemOpen
  \bibfield  {author} {\bibinfo {author} {\bibfnamefont {Y.-F.}\ \bibnamefont {Wang}}, \bibinfo {author} {\bibfnamefont {H.}~\bibnamefont {Yao}}, \bibinfo {author} {\bibfnamefont {C.-D.}\ \bibnamefont {Gong}},\ and\ \bibinfo {author} {\bibfnamefont {D.~N.}\ \bibnamefont {Sheng}},\ }\bibfield  {title} {\bibinfo {title} {Fractional quantum hall effect in topological flat bands with chern number two},\ }\href {https://doi.org/10.1103/PhysRevB.86.201101} {\bibfield  {journal} {\bibinfo  {journal} {Phys. Rev. B}\ }\textbf {\bibinfo {volume} {86}},\ \bibinfo {pages} {201101} (\bibinfo {year} {2012}{\natexlab{b}})}\BibitemShut {NoStop}%
\bibitem [{\citenamefont {Liu}\ \emph {et~al.}(2012{\natexlab{a}})\citenamefont {Liu}, \citenamefont {Bergholtz}, \citenamefont {Fan},\ and\ \citenamefont {L\"auchli}}]{ZLiu1}%
  \BibitemOpen
  \bibfield  {author} {\bibinfo {author} {\bibfnamefont {Z.}~\bibnamefont {Liu}}, \bibinfo {author} {\bibfnamefont {E.~J.}\ \bibnamefont {Bergholtz}}, \bibinfo {author} {\bibfnamefont {H.}~\bibnamefont {Fan}},\ and\ \bibinfo {author} {\bibfnamefont {A.~M.}\ \bibnamefont {L\"auchli}},\ }\bibfield  {title} {\bibinfo {title} {Fractional chern insulators in topological flat bands with higher chern number},\ }\href {https://doi.org/10.1103/PhysRevLett.109.186805} {\bibfield  {journal} {\bibinfo  {journal} {Phys. Rev. Lett.}\ }\textbf {\bibinfo {volume} {109}},\ \bibinfo {pages} {186805} (\bibinfo {year} {2012}{\natexlab{a}})}\BibitemShut {NoStop}%
\bibitem [{\citenamefont {Scaffidi}\ and\ \citenamefont {M\"oller}(2012)}]{Scaffidi}%
  \BibitemOpen
  \bibfield  {author} {\bibinfo {author} {\bibfnamefont {T.}~\bibnamefont {Scaffidi}}\ and\ \bibinfo {author} {\bibfnamefont {G.}~\bibnamefont {M\"oller}},\ }\bibfield  {title} {\bibinfo {title} {Adiabatic continuation of fractional chern insulators to fractional quantum hall states},\ }\href {https://doi.org/10.1103/PhysRevLett.109.246805} {\bibfield  {journal} {\bibinfo  {journal} {Phys. Rev. Lett.}\ }\textbf {\bibinfo {volume} {109}},\ \bibinfo {pages} {246805} (\bibinfo {year} {2012})}\BibitemShut {NoStop}%
\bibitem [{\citenamefont {Wu}\ \emph {et~al.}(2013)\citenamefont {Wu}, \citenamefont {Regnault},\ and\ \citenamefont {Bernevig}}]{YLWu0}%
  \BibitemOpen
  \bibfield  {author} {\bibinfo {author} {\bibfnamefont {Y.-L.}\ \bibnamefont {Wu}}, \bibinfo {author} {\bibfnamefont {N.}~\bibnamefont {Regnault}},\ and\ \bibinfo {author} {\bibfnamefont {B.~A.}\ \bibnamefont {Bernevig}},\ }\bibfield  {title} {\bibinfo {title} {Bloch model wave functions and pseudopotentials for all fractional chern insulators},\ }\href {https://doi.org/10.1103/PhysRevLett.110.106802} {\bibfield  {journal} {\bibinfo  {journal} {Phys. Rev. Lett.}\ }\textbf {\bibinfo {volume} {110}},\ \bibinfo {pages} {106802} (\bibinfo {year} {2013})}\BibitemShut {NoStop}%
\bibitem [{\citenamefont {Lee}\ \emph {et~al.}(2013)\citenamefont {Lee}, \citenamefont {Thomale},\ and\ \citenamefont {Qi}}]{Ronny0}%
  \BibitemOpen
  \bibfield  {author} {\bibinfo {author} {\bibfnamefont {C.~H.}\ \bibnamefont {Lee}}, \bibinfo {author} {\bibfnamefont {R.}~\bibnamefont {Thomale}},\ and\ \bibinfo {author} {\bibfnamefont {X.-L.}\ \bibnamefont {Qi}},\ }\bibfield  {title} {\bibinfo {title} {Pseudopotential formalism for fractional chern insulators},\ }\href {https://doi.org/10.1103/PhysRevB.88.035101} {\bibfield  {journal} {\bibinfo  {journal} {Phys. Rev. B}\ }\textbf {\bibinfo {volume} {88}},\ \bibinfo {pages} {035101} (\bibinfo {year} {2013})}\BibitemShut {NoStop}%
\bibitem [{\citenamefont {Luo}\ \emph {et~al.}(2013)\citenamefont {Luo}, \citenamefont {Chen}, \citenamefont {Wang},\ and\ \citenamefont {Gong}}]{YFW4}%
  \BibitemOpen
  \bibfield  {author} {\bibinfo {author} {\bibfnamefont {W.-W.}\ \bibnamefont {Luo}}, \bibinfo {author} {\bibfnamefont {W.-C.}\ \bibnamefont {Chen}}, \bibinfo {author} {\bibfnamefont {Y.-F.}\ \bibnamefont {Wang}},\ and\ \bibinfo {author} {\bibfnamefont {C.-D.}\ \bibnamefont {Gong}},\ }\bibfield  {title} {\bibinfo {title} {Edge excitations in fractional chern insulators},\ }\href {https://doi.org/10.1103/PhysRevB.88.161109} {\bibfield  {journal} {\bibinfo  {journal} {Phys. Rev. B}\ }\textbf {\bibinfo {volume} {88}},\ \bibinfo {pages} {161109} (\bibinfo {year} {2013})}\BibitemShut {NoStop}%
\bibitem [{\citenamefont {Liu}\ \emph {et~al.}(2013{\natexlab{a}})\citenamefont {Liu}, \citenamefont {Kovrizhin},\ and\ \citenamefont {Bergholtz}}]{ZLiu2}%
  \BibitemOpen
  \bibfield  {author} {\bibinfo {author} {\bibfnamefont {Z.}~\bibnamefont {Liu}}, \bibinfo {author} {\bibfnamefont {D.~L.}\ \bibnamefont {Kovrizhin}},\ and\ \bibinfo {author} {\bibfnamefont {E.~J.}\ \bibnamefont {Bergholtz}},\ }\bibfield  {title} {\bibinfo {title} {Bulk-edge correspondence in fractional chern insulators},\ }\href {https://doi.org/10.1103/PhysRevB.88.081106} {\bibfield  {journal} {\bibinfo  {journal} {Phys. Rev. B}\ }\textbf {\bibinfo {volume} {88}},\ \bibinfo {pages} {081106} (\bibinfo {year} {2013}{\natexlab{a}})}\BibitemShut {NoStop}%
\bibitem [{\citenamefont {Liu}\ \emph {et~al.}(2013{\natexlab{b}})\citenamefont {Liu}, \citenamefont {Bergholtz},\ and\ \citenamefont {Kapit}}]{ZLiu3}%
  \BibitemOpen
  \bibfield  {author} {\bibinfo {author} {\bibfnamefont {Z.}~\bibnamefont {Liu}}, \bibinfo {author} {\bibfnamefont {E.~J.}\ \bibnamefont {Bergholtz}},\ and\ \bibinfo {author} {\bibfnamefont {E.}~\bibnamefont {Kapit}},\ }\bibfield  {title} {\bibinfo {title} {Non-abelian fractional chern insulators from long-range interactions},\ }\href {https://doi.org/10.1103/PhysRevB.88.205101} {\bibfield  {journal} {\bibinfo  {journal} {Phys. Rev. B}\ }\textbf {\bibinfo {volume} {88}},\ \bibinfo {pages} {205101} (\bibinfo {year} {2013}{\natexlab{b}})}\BibitemShut {NoStop}%
\bibitem [{\citenamefont {Claassen}\ \emph {et~al.}(2015)\citenamefont {Claassen}, \citenamefont {Lee}, \citenamefont {Thomale}, \citenamefont {Qi},\ and\ \citenamefont {Devereaux}}]{Ronny1}%
  \BibitemOpen
  \bibfield  {author} {\bibinfo {author} {\bibfnamefont {M.}~\bibnamefont {Claassen}}, \bibinfo {author} {\bibfnamefont {C.~H.}\ \bibnamefont {Lee}}, \bibinfo {author} {\bibfnamefont {R.}~\bibnamefont {Thomale}}, \bibinfo {author} {\bibfnamefont {X.-L.}\ \bibnamefont {Qi}},\ and\ \bibinfo {author} {\bibfnamefont {T.~P.}\ \bibnamefont {Devereaux}},\ }\bibfield  {title} {\bibinfo {title} {Position-momentum duality and fractional quantum hall effect in chern insulators},\ }\href {https://doi.org/10.1103/PhysRevLett.114.236802} {\bibfield  {journal} {\bibinfo  {journal} {Phys. Rev. Lett.}\ }\textbf {\bibinfo {volume} {114}},\ \bibinfo {pages} {236802} (\bibinfo {year} {2015})}\BibitemShut {NoStop}%
\bibitem [{\citenamefont {Parameswaran}\ \emph {et~al.}(2013)\citenamefont {Parameswaran}, \citenamefont {Roy},\ and\ \citenamefont {Sondhi}}]{FCI_reviews}%
  \BibitemOpen
  \bibfield  {author} {\bibinfo {author} {\bibfnamefont {S.~A.}\ \bibnamefont {Parameswaran}}, \bibinfo {author} {\bibfnamefont {R.}~\bibnamefont {Roy}},\ and\ \bibinfo {author} {\bibfnamefont {S.~L.}\ \bibnamefont {Sondhi}},\ }\bibfield  {title} {\bibinfo {title} {Fractional quantum hall physics in topological flat bands},\ }\href {https://doi.org/https://doi.org/10.1016/j.crhy.2013.04.003} {\bibfield  {journal} {\bibinfo  {journal} {Comptes Rendus Physique}\ }\textbf {\bibinfo {volume} {14}},\ \bibinfo {pages} {816 } (\bibinfo {year} {2013})}\BibitemShut {NoStop}%
\bibitem [{\citenamefont {Bergholtz}\ and\ \citenamefont {Liu}(2013)}]{FCI_reviews1}%
  \BibitemOpen
  \bibfield  {author} {\bibinfo {author} {\bibfnamefont {E.~J.}\ \bibnamefont {Bergholtz}}\ and\ \bibinfo {author} {\bibfnamefont {Z.}~\bibnamefont {Liu}},\ }\bibfield  {title} {\bibinfo {title} {Topological flat band models and fractional chern insulators},\ }\href {https://doi.org/10.1142/S021797921330017X} {\bibfield  {journal} {\bibinfo  {journal} {International Journal of Modern Physics B}\ }\textbf {\bibinfo {volume} {27}},\ \bibinfo {pages} {1330017} (\bibinfo {year} {2013})}\BibitemShut {NoStop}%
\bibitem [{\citenamefont {Liu}\ and\ \citenamefont {Bergholtz}(2023)}]{FCI_reviews2}%
  \BibitemOpen
  \bibfield  {author} {\bibinfo {author} {\bibfnamefont {Z.}~\bibnamefont {Liu}}\ and\ \bibinfo {author} {\bibfnamefont {E.~J.}\ \bibnamefont {Bergholtz}},\ }\bibfield  {title} {\bibinfo {title} {Recent developments in fractional chern insulators},\ }in\ \href {https://doi.org/https://doi.org/10.1016/B978-0-323-90800-9.00136-0} {\emph {\bibinfo {booktitle} {Reference Module in Materials Science and Materials Engineering}}}\ (\bibinfo  {publisher} {Elsevier},\ \bibinfo {year} {2023})\BibitemShut {NoStop}%
\bibitem [{\citenamefont {Cai}\ \emph {et~al.}(2023)\citenamefont {Cai}, \citenamefont {Anderson}, \citenamefont {Wang}, \citenamefont {Zhang}, \citenamefont {Liu}, \citenamefont {Holtzmann}, \citenamefont {Zhang}, \citenamefont {Fan}, \citenamefont {Taniguchi}, \citenamefont {Watanabe}, \citenamefont {Ran}, \citenamefont {Cao}, \citenamefont {Fu}, \citenamefont {Xiao}, \citenamefont {Yao},\ and\ \citenamefont {Xu}}]{Cai2023}%
  \BibitemOpen
  \bibfield  {author} {\bibinfo {author} {\bibfnamefont {J.}~\bibnamefont {Cai}}, \bibinfo {author} {\bibfnamefont {E.}~\bibnamefont {Anderson}}, \bibinfo {author} {\bibfnamefont {C.}~\bibnamefont {Wang}}, \bibinfo {author} {\bibfnamefont {X.}~\bibnamefont {Zhang}}, \bibinfo {author} {\bibfnamefont {X.}~\bibnamefont {Liu}}, \bibinfo {author} {\bibfnamefont {W.}~\bibnamefont {Holtzmann}}, \bibinfo {author} {\bibfnamefont {Y.}~\bibnamefont {Zhang}}, \bibinfo {author} {\bibfnamefont {F.}~\bibnamefont {Fan}}, \bibinfo {author} {\bibfnamefont {T.}~\bibnamefont {Taniguchi}}, \bibinfo {author} {\bibfnamefont {K.}~\bibnamefont {Watanabe}}, \bibinfo {author} {\bibfnamefont {Y.}~\bibnamefont {Ran}}, \bibinfo {author} {\bibfnamefont {T.}~\bibnamefont {Cao}}, \bibinfo {author} {\bibfnamefont {L.}~\bibnamefont {Fu}}, \bibinfo {author} {\bibfnamefont {D.}~\bibnamefont {Xiao}}, \bibinfo {author} {\bibfnamefont {W.}~\bibnamefont {Yao}},\ and\ \bibinfo {author} {\bibfnamefont {X.}~\bibnamefont {Xu}},\ }\bibfield  {title}
  {\bibinfo {title} {Signatures of fractional quantum anomalous hall states in twisted mote2},\ }\href {https://doi.org/10.1038/s41586-023-06289-w} {\bibfield  {journal} {\bibinfo  {journal} {Nature}\ }\textbf {\bibinfo {volume} {622}},\ \bibinfo {pages} {63} (\bibinfo {year} {2023})}\BibitemShut {NoStop}%
\bibitem [{\citenamefont {Zeng}\ \emph {et~al.}(2023)\citenamefont {Zeng}, \citenamefont {Xia}, \citenamefont {Kang}, \citenamefont {Zhu}, \citenamefont {Kn{\"u}ppel}, \citenamefont {Vaswani}, \citenamefont {Watanabe}, \citenamefont {Taniguchi}, \citenamefont {Mak},\ and\ \citenamefont {Shan}}]{Zeng2023}%
  \BibitemOpen
  \bibfield  {author} {\bibinfo {author} {\bibfnamefont {Y.}~\bibnamefont {Zeng}}, \bibinfo {author} {\bibfnamefont {Z.}~\bibnamefont {Xia}}, \bibinfo {author} {\bibfnamefont {K.}~\bibnamefont {Kang}}, \bibinfo {author} {\bibfnamefont {J.}~\bibnamefont {Zhu}}, \bibinfo {author} {\bibfnamefont {P.}~\bibnamefont {Kn{\"u}ppel}}, \bibinfo {author} {\bibfnamefont {C.}~\bibnamefont {Vaswani}}, \bibinfo {author} {\bibfnamefont {K.}~\bibnamefont {Watanabe}}, \bibinfo {author} {\bibfnamefont {T.}~\bibnamefont {Taniguchi}}, \bibinfo {author} {\bibfnamefont {K.~F.}\ \bibnamefont {Mak}},\ and\ \bibinfo {author} {\bibfnamefont {J.}~\bibnamefont {Shan}},\ }\bibfield  {title} {\bibinfo {title} {Thermodynamic evidence of fractional chern insulator in moir{\'e} mote2},\ }\href {https://doi.org/10.1038/s41586-023-06452-3} {\bibfield  {journal} {\bibinfo  {journal} {Nature}\ }\textbf {\bibinfo {volume} {622}},\ \bibinfo {pages} {69} (\bibinfo {year} {2023})}\BibitemShut {NoStop}%
\bibitem [{\citenamefont {Park}\ \emph {et~al.}(2023)\citenamefont {Park}, \citenamefont {Cai}, \citenamefont {Anderson}, \citenamefont {Zhang}, \citenamefont {Zhu}, \citenamefont {Liu}, \citenamefont {Wang}, \citenamefont {Holtzmann}, \citenamefont {Hu}, \citenamefont {Liu}, \citenamefont {Taniguchi}, \citenamefont {Watanabe}, \citenamefont {Chu}, \citenamefont {Cao}, \citenamefont {Fu}, \citenamefont {Yao}, \citenamefont {Chang}, \citenamefont {Cobden}, \citenamefont {Xiao},\ and\ \citenamefont {Xu}}]{Park2023}%
  \BibitemOpen
  \bibfield  {author} {\bibinfo {author} {\bibfnamefont {H.}~\bibnamefont {Park}}, \bibinfo {author} {\bibfnamefont {J.}~\bibnamefont {Cai}}, \bibinfo {author} {\bibfnamefont {E.}~\bibnamefont {Anderson}}, \bibinfo {author} {\bibfnamefont {Y.}~\bibnamefont {Zhang}}, \bibinfo {author} {\bibfnamefont {J.}~\bibnamefont {Zhu}}, \bibinfo {author} {\bibfnamefont {X.}~\bibnamefont {Liu}}, \bibinfo {author} {\bibfnamefont {C.}~\bibnamefont {Wang}}, \bibinfo {author} {\bibfnamefont {W.}~\bibnamefont {Holtzmann}}, \bibinfo {author} {\bibfnamefont {C.}~\bibnamefont {Hu}}, \bibinfo {author} {\bibfnamefont {Z.}~\bibnamefont {Liu}}, \bibinfo {author} {\bibfnamefont {T.}~\bibnamefont {Taniguchi}}, \bibinfo {author} {\bibfnamefont {K.}~\bibnamefont {Watanabe}}, \bibinfo {author} {\bibfnamefont {J.-H.}\ \bibnamefont {Chu}}, \bibinfo {author} {\bibfnamefont {T.}~\bibnamefont {Cao}}, \bibinfo {author} {\bibfnamefont {L.}~\bibnamefont {Fu}}, \bibinfo {author} {\bibfnamefont {W.}~\bibnamefont {Yao}}, \bibinfo {author}
  {\bibfnamefont {C.-Z.}\ \bibnamefont {Chang}}, \bibinfo {author} {\bibfnamefont {D.}~\bibnamefont {Cobden}}, \bibinfo {author} {\bibfnamefont {D.}~\bibnamefont {Xiao}},\ and\ \bibinfo {author} {\bibfnamefont {X.}~\bibnamefont {Xu}},\ }\bibfield  {title} {\bibinfo {title} {Observation of fractionally quantized anomalous hall effect},\ }\href {https://doi.org/10.1038/s41586-023-06536-0} {\bibfield  {journal} {\bibinfo  {journal} {Nature}\ }\textbf {\bibinfo {volume} {622}},\ \bibinfo {pages} {74} (\bibinfo {year} {2023})}\BibitemShut {NoStop}%
\bibitem [{\citenamefont {Xu}\ \emph {et~al.}(2023)\citenamefont {Xu}, \citenamefont {Sun}, \citenamefont {Jia}, \citenamefont {Liu}, \citenamefont {Xu}, \citenamefont {Li}, \citenamefont {Gu}, \citenamefont {Watanabe}, \citenamefont {Taniguchi}, \citenamefont {Tong}, \citenamefont {Jia}, \citenamefont {Shi}, \citenamefont {Jiang}, \citenamefont {Zhang}, \citenamefont {Liu},\ and\ \citenamefont {Li}}]{XuFan2023}%
  \BibitemOpen
  \bibfield  {author} {\bibinfo {author} {\bibfnamefont {F.}~\bibnamefont {Xu}}, \bibinfo {author} {\bibfnamefont {Z.}~\bibnamefont {Sun}}, \bibinfo {author} {\bibfnamefont {T.}~\bibnamefont {Jia}}, \bibinfo {author} {\bibfnamefont {C.}~\bibnamefont {Liu}}, \bibinfo {author} {\bibfnamefont {C.}~\bibnamefont {Xu}}, \bibinfo {author} {\bibfnamefont {C.}~\bibnamefont {Li}}, \bibinfo {author} {\bibfnamefont {Y.}~\bibnamefont {Gu}}, \bibinfo {author} {\bibfnamefont {K.}~\bibnamefont {Watanabe}}, \bibinfo {author} {\bibfnamefont {T.}~\bibnamefont {Taniguchi}}, \bibinfo {author} {\bibfnamefont {B.}~\bibnamefont {Tong}}, \bibinfo {author} {\bibfnamefont {J.}~\bibnamefont {Jia}}, \bibinfo {author} {\bibfnamefont {Z.}~\bibnamefont {Shi}}, \bibinfo {author} {\bibfnamefont {S.}~\bibnamefont {Jiang}}, \bibinfo {author} {\bibfnamefont {Y.}~\bibnamefont {Zhang}}, \bibinfo {author} {\bibfnamefont {X.}~\bibnamefont {Liu}},\ and\ \bibinfo {author} {\bibfnamefont {T.}~\bibnamefont {Li}},\ }\bibfield  {title} {\bibinfo {title}
  {Observation of integer and fractional quantum anomalous hall effects in twisted bilayer ${\mathrm{mote}}_{2}$},\ }\href {https://doi.org/10.1103/PhysRevX.13.031037} {\bibfield  {journal} {\bibinfo  {journal} {Phys. Rev. X}\ }\textbf {\bibinfo {volume} {13}},\ \bibinfo {pages} {031037} (\bibinfo {year} {2023})}\BibitemShut {NoStop}%
\bibitem [{\citenamefont {Lu}\ \emph {et~al.}(2024)\citenamefont {Lu}, \citenamefont {Han}, \citenamefont {Yao}, \citenamefont {Reddy}, \citenamefont {Yang}, \citenamefont {Seo}, \citenamefont {Watanabe}, \citenamefont {Taniguchi}, \citenamefont {Fu},\ and\ \citenamefont {Ju}}]{lu2023fractional}%
  \BibitemOpen
  \bibfield  {author} {\bibinfo {author} {\bibfnamefont {Z.}~\bibnamefont {Lu}}, \bibinfo {author} {\bibfnamefont {T.}~\bibnamefont {Han}}, \bibinfo {author} {\bibfnamefont {Y.}~\bibnamefont {Yao}}, \bibinfo {author} {\bibfnamefont {A.~P.}\ \bibnamefont {Reddy}}, \bibinfo {author} {\bibfnamefont {J.}~\bibnamefont {Yang}}, \bibinfo {author} {\bibfnamefont {J.}~\bibnamefont {Seo}}, \bibinfo {author} {\bibfnamefont {K.}~\bibnamefont {Watanabe}}, \bibinfo {author} {\bibfnamefont {T.}~\bibnamefont {Taniguchi}}, \bibinfo {author} {\bibfnamefont {L.}~\bibnamefont {Fu}},\ and\ \bibinfo {author} {\bibfnamefont {L.}~\bibnamefont {Ju}},\ }\bibfield  {title} {\bibinfo {title} {Fractional quantum anomalous hall effect in multilayer graphene},\ }\href {https://doi.org/10.1038/s41586-023-07010-7} {\bibfield  {journal} {\bibinfo  {journal} {Nature}\ }\textbf {\bibinfo {volume} {626}},\ \bibinfo {pages} {759} (\bibinfo {year} {2024})}\BibitemShut {NoStop}%
\bibitem [{\citenamefont {Sun}\ \emph {et~al.}(2011)\citenamefont {Sun}, \citenamefont {Gu}, \citenamefont {Katsura},\ and\ \citenamefont {Das~Sarma}}]{TFBCB}%
  \BibitemOpen
  \bibfield  {author} {\bibinfo {author} {\bibfnamefont {K.}~\bibnamefont {Sun}}, \bibinfo {author} {\bibfnamefont {Z.}~\bibnamefont {Gu}}, \bibinfo {author} {\bibfnamefont {H.}~\bibnamefont {Katsura}},\ and\ \bibinfo {author} {\bibfnamefont {S.}~\bibnamefont {Das~Sarma}},\ }\bibfield  {title} {\bibinfo {title} {Nearly flatbands with nontrivial topology},\ }\href {https://doi.org/10.1103/PhysRevLett.106.236803} {\bibfield  {journal} {\bibinfo  {journal} {Phys. Rev. Lett.}\ }\textbf {\bibinfo {volume} {106}},\ \bibinfo {pages} {236803} (\bibinfo {year} {2011})}\BibitemShut {NoStop}%
\bibitem [{\citenamefont {Kapit}\ and\ \citenamefont {Mueller}(2010)}]{TFBKapit}%
  \BibitemOpen
  \bibfield  {author} {\bibinfo {author} {\bibfnamefont {E.}~\bibnamefont {Kapit}}\ and\ \bibinfo {author} {\bibfnamefont {E.}~\bibnamefont {Mueller}},\ }\bibfield  {title} {\bibinfo {title} {Exact parent hamiltonian for the quantum hall states in a lattice},\ }\href {https://doi.org/10.1103/PhysRevLett.105.215303} {\bibfield  {journal} {\bibinfo  {journal} {Phys. Rev. Lett.}\ }\textbf {\bibinfo {volume} {105}},\ \bibinfo {pages} {215303} (\bibinfo {year} {2010})}\BibitemShut {NoStop}%
\bibitem [{\citenamefont {Hu}\ \emph {et~al.}(2011)\citenamefont {Hu}, \citenamefont {Kargarian},\ and\ \citenamefont {Fiete}}]{TFBRuby}%
  \BibitemOpen
  \bibfield  {author} {\bibinfo {author} {\bibfnamefont {X.}~\bibnamefont {Hu}}, \bibinfo {author} {\bibfnamefont {M.}~\bibnamefont {Kargarian}},\ and\ \bibinfo {author} {\bibfnamefont {G.~A.}\ \bibnamefont {Fiete}},\ }\bibfield  {title} {\bibinfo {title} {Topological insulators and fractional quantum hall effect on the ruby lattice},\ }\href {https://doi.org/10.1103/PhysRevB.84.155116} {\bibfield  {journal} {\bibinfo  {journal} {Phys. Rev. B}\ }\textbf {\bibinfo {volume} {84}},\ \bibinfo {pages} {155116} (\bibinfo {year} {2011})}\BibitemShut {NoStop}%
\bibitem [{\citenamefont {Chen}\ \emph {et~al.}(2012)\citenamefont {Chen}, \citenamefont {Liu}, \citenamefont {Wang},\ and\ \citenamefont {Gong}}]{TFBStar}%
  \BibitemOpen
  \bibfield  {author} {\bibinfo {author} {\bibfnamefont {W.-C.}\ \bibnamefont {Chen}}, \bibinfo {author} {\bibfnamefont {R.}~\bibnamefont {Liu}}, \bibinfo {author} {\bibfnamefont {Y.-F.}\ \bibnamefont {Wang}},\ and\ \bibinfo {author} {\bibfnamefont {C.-D.}\ \bibnamefont {Gong}},\ }\bibfield  {title} {\bibinfo {title} {Topological quantum phase transitions and topological flat bands on the star lattice},\ }\href {https://doi.org/10.1103/PhysRevB.86.085311} {\bibfield  {journal} {\bibinfo  {journal} {Phys. Rev. B}\ }\textbf {\bibinfo {volume} {86}},\ \bibinfo {pages} {085311} (\bibinfo {year} {2012})}\BibitemShut {NoStop}%
\bibitem [{\citenamefont {Wang}\ and\ \citenamefont {Ran}(2011)}]{TFBdice}%
  \BibitemOpen
  \bibfield  {author} {\bibinfo {author} {\bibfnamefont {F.}~\bibnamefont {Wang}}\ and\ \bibinfo {author} {\bibfnamefont {Y.}~\bibnamefont {Ran}},\ }\bibfield  {title} {\bibinfo {title} {Nearly flat band with chern number $c=2$ on the dice lattice},\ }\href {https://doi.org/10.1103/PhysRevB.84.241103} {\bibfield  {journal} {\bibinfo  {journal} {Phys. Rev. B}\ }\textbf {\bibinfo {volume} {84}},\ \bibinfo {pages} {241103} (\bibinfo {year} {2011})}\BibitemShut {NoStop}%
\bibitem [{\citenamefont {Liu}\ \emph {et~al.}(2012{\natexlab{b}})\citenamefont {Liu}, \citenamefont {Chen}, \citenamefont {Wang},\ and\ \citenamefont {Gong}}]{TFBKG1}%
  \BibitemOpen
  \bibfield  {author} {\bibinfo {author} {\bibfnamefont {R.}~\bibnamefont {Liu}}, \bibinfo {author} {\bibfnamefont {W.-C.}\ \bibnamefont {Chen}}, \bibinfo {author} {\bibfnamefont {Y.-F.}\ \bibnamefont {Wang}},\ and\ \bibinfo {author} {\bibfnamefont {C.-D.}\ \bibnamefont {Gong}},\ }\bibfield  {title} {\bibinfo {title} {Topological quantum phase transitions and topological flat bands on the kagom{\'{e}} lattice},\ }\href {https://doi.org/10.1088/0953-8984/24/30/305602} {\bibfield  {journal} {\bibinfo  {journal} {Journal of Physics: Condensed Matter}\ }\textbf {\bibinfo {volume} {24}},\ \bibinfo {pages} {305602} (\bibinfo {year} {2012}{\natexlab{b}})}\BibitemShut {NoStop}%
\bibitem [{\citenamefont {Liu}\ \emph {et~al.}(2013{\natexlab{c}})\citenamefont {Liu}, \citenamefont {Chen}, \citenamefont {Wang},\ and\ \citenamefont {Gong}}]{TFBSQOC}%
  \BibitemOpen
  \bibfield  {author} {\bibinfo {author} {\bibfnamefont {X.-P.}\ \bibnamefont {Liu}}, \bibinfo {author} {\bibfnamefont {W.-C.}\ \bibnamefont {Chen}}, \bibinfo {author} {\bibfnamefont {Y.-F.}\ \bibnamefont {Wang}},\ and\ \bibinfo {author} {\bibfnamefont {C.-D.}\ \bibnamefont {Gong}},\ }\bibfield  {title} {\bibinfo {title} {Topological quantum phase transitions on the kagom{\'{e}} and square{\textendash}octagon lattices},\ }\href {https://doi.org/10.1088/0953-8984/25/30/305602} {\bibfield  {journal} {\bibinfo  {journal} {Journal of Physics: Condensed Matter}\ }\textbf {\bibinfo {volume} {25}},\ \bibinfo {pages} {305602} (\bibinfo {year} {2013}{\natexlab{c}})}\BibitemShut {NoStop}%
\bibitem [{\citenamefont {Lan}\ \emph {et~al.}(2023)\citenamefont {Lan}, \citenamefont {He},\ and\ \citenamefont {Wang}}]{Lan}%
  \BibitemOpen
  \bibfield  {author} {\bibinfo {author} {\bibfnamefont {Z.-Y.}\ \bibnamefont {Lan}}, \bibinfo {author} {\bibfnamefont {A.-L.}\ \bibnamefont {He}},\ and\ \bibinfo {author} {\bibfnamefont {Y.-F.}\ \bibnamefont {Wang}},\ }\bibfield  {title} {\bibinfo {title} {Flat bands with high chern numbers and multiple flat bands in multifold staggered-flux models},\ }\href {https://doi.org/10.1103/PhysRevB.107.235116} {\bibfield  {journal} {\bibinfo  {journal} {Phys. Rev. B}\ }\textbf {\bibinfo {volume} {107}},\ \bibinfo {pages} {235116} (\bibinfo {year} {2023})}\BibitemShut {NoStop}%
\bibitem [{\citenamefont {Zhang}\ and\ \citenamefont {Shi}(2016)}]{YHZhang}%
  \BibitemOpen
  \bibfield  {author} {\bibinfo {author} {\bibfnamefont {Y.}~\bibnamefont {Zhang}}\ and\ \bibinfo {author} {\bibfnamefont {J.}~\bibnamefont {Shi}},\ }\bibfield  {title} {\bibinfo {title} {Mapping a fractional quantum hall state to a fractional chern insulator},\ }\href {https://doi.org/10.1103/PhysRevB.93.165129} {\bibfield  {journal} {\bibinfo  {journal} {Phys. Rev. B}\ }\textbf {\bibinfo {volume} {93}},\ \bibinfo {pages} {165129} (\bibinfo {year} {2016})}\BibitemShut {NoStop}%
\bibitem [{\citenamefont {Laughlin}(1983)}]{Laughlin1983}%
  \BibitemOpen
  \bibfield  {author} {\bibinfo {author} {\bibfnamefont {R.~B.}\ \bibnamefont {Laughlin}},\ }\bibfield  {title} {\bibinfo {title} {Anomalous quantum hall effect: An incompressible quantum fluid with fractionally charged excitations},\ }\href {https://doi.org/10.1103/PhysRevLett.50.1395} {\bibfield  {journal} {\bibinfo  {journal} {Phys. Rev. Lett.}\ }\textbf {\bibinfo {volume} {50}},\ \bibinfo {pages} {1395} (\bibinfo {year} {1983})}\BibitemShut {NoStop}%
\bibitem [{\citenamefont {Haldane}(1991)}]{GPP1}%
  \BibitemOpen
  \bibfield  {author} {\bibinfo {author} {\bibfnamefont {F.~D.~M.}\ \bibnamefont {Haldane}},\ }\bibfield  {title} {\bibinfo {title} {``fractional statistics'' in arbitrary dimensions: A generalization of the pauli principle},\ }\href {https://doi.org/10.1103/PhysRevLett.67.937} {\bibfield  {journal} {\bibinfo  {journal} {Phys. Rev. Lett.}\ }\textbf {\bibinfo {volume} {67}},\ \bibinfo {pages} {937} (\bibinfo {year} {1991})}\BibitemShut {NoStop}%
\bibitem [{\citenamefont {Wu}(1994)}]{GPP2}%
  \BibitemOpen
  \bibfield  {author} {\bibinfo {author} {\bibfnamefont {Y.-S.}\ \bibnamefont {Wu}},\ }\bibfield  {title} {\bibinfo {title} {Statistical distribution for generalized ideal gas of fractional-statistics particles},\ }\href {https://doi.org/10.1103/PhysRevLett.73.922} {\bibfield  {journal} {\bibinfo  {journal} {Phys. Rev. Lett.}\ }\textbf {\bibinfo {volume} {73}},\ \bibinfo {pages} {922} (\bibinfo {year} {1994})}\BibitemShut {NoStop}%
\bibitem [{\citenamefont {Bergholtz}\ and\ \citenamefont {Karlhede}(2008)}]{GPP3}%
  \BibitemOpen
  \bibfield  {author} {\bibinfo {author} {\bibfnamefont {E.~J.}\ \bibnamefont {Bergholtz}}\ and\ \bibinfo {author} {\bibfnamefont {A.}~\bibnamefont {Karlhede}},\ }\bibfield  {title} {\bibinfo {title} {Quantum hall system in tao-thouless limit},\ }\href {https://doi.org/10.1103/PhysRevB.77.155308} {\bibfield  {journal} {\bibinfo  {journal} {Phys. Rev. B}\ }\textbf {\bibinfo {volume} {77}},\ \bibinfo {pages} {155308} (\bibinfo {year} {2008})}\BibitemShut {NoStop}%
\bibitem [{\citenamefont {He}\ \emph {et~al.}(2015)\citenamefont {He}, \citenamefont {Luo}, \citenamefont {Wang},\ and\ \citenamefont {Gong}}]{HeAL1}%
  \BibitemOpen
  \bibfield  {author} {\bibinfo {author} {\bibfnamefont {A.-L.}\ \bibnamefont {He}}, \bibinfo {author} {\bibfnamefont {W.-W.}\ \bibnamefont {Luo}}, \bibinfo {author} {\bibfnamefont {Y.-F.}\ \bibnamefont {Wang}},\ and\ \bibinfo {author} {\bibfnamefont {C.-D.}\ \bibnamefont {Gong}},\ }\bibfield  {title} {\bibinfo {title} {Wave functions for fractional chern insulators in disk geometry},\ }\href {https://doi.org/10.1088/1367-2630/17/12/125005} {\bibfield  {journal} {\bibinfo  {journal} {New Journal of Physics}\ }\textbf {\bibinfo {volume} {17}},\ \bibinfo {pages} {125005} (\bibinfo {year} {2015})}\BibitemShut {NoStop}%
\bibitem [{\citenamefont {He}\ \emph {et~al.}(2020)\citenamefont {He}, \citenamefont {Luo}, \citenamefont {Yao},\ and\ \citenamefont {Wang}}]{HeAL2}%
  \BibitemOpen
  \bibfield  {author} {\bibinfo {author} {\bibfnamefont {A.-L.}\ \bibnamefont {He}}, \bibinfo {author} {\bibfnamefont {W.-W.}\ \bibnamefont {Luo}}, \bibinfo {author} {\bibfnamefont {H.}~\bibnamefont {Yao}},\ and\ \bibinfo {author} {\bibfnamefont {Y.-F.}\ \bibnamefont {Wang}},\ }\bibfield  {title} {\bibinfo {title} {Non-abelian fractional chern insulator in disk geometry},\ }\href {https://doi.org/10.1103/PhysRevB.101.165127} {\bibfield  {journal} {\bibinfo  {journal} {Phys. Rev. B}\ }\textbf {\bibinfo {volume} {101}},\ \bibinfo {pages} {165127} (\bibinfo {year} {2020})}\BibitemShut {NoStop}%
\bibitem [{\citenamefont {Bernevig}\ and\ \citenamefont {Haldane}(2008{\natexlab{a}})}]{Bernevig1}%
  \BibitemOpen
  \bibfield  {author} {\bibinfo {author} {\bibfnamefont {B.~A.}\ \bibnamefont {Bernevig}}\ and\ \bibinfo {author} {\bibfnamefont {F.~D.~M.}\ \bibnamefont {Haldane}},\ }\bibfield  {title} {\bibinfo {title} {Model fractional quantum hall states and jack polynomials},\ }\href {https://doi.org/10.1103/PhysRevLett.100.246802} {\bibfield  {journal} {\bibinfo  {journal} {Phys. Rev. Lett.}\ }\textbf {\bibinfo {volume} {100}},\ \bibinfo {pages} {246802} (\bibinfo {year} {2008}{\natexlab{a}})}\BibitemShut {NoStop}%
\bibitem [{\citenamefont {Bernevig}\ and\ \citenamefont {Haldane}(2008{\natexlab{b}})}]{Bernevig2}%
  \BibitemOpen
  \bibfield  {author} {\bibinfo {author} {\bibfnamefont {B.~A.}\ \bibnamefont {Bernevig}}\ and\ \bibinfo {author} {\bibfnamefont {F.~D.~M.}\ \bibnamefont {Haldane}},\ }\bibfield  {title} {\bibinfo {title} {Properties of non-abelian fractional quantum hall states at filling $\ensuremath{\nu}=k/r$},\ }\href {https://doi.org/10.1103/PhysRevLett.101.246806} {\bibfield  {journal} {\bibinfo  {journal} {Phys. Rev. Lett.}\ }\textbf {\bibinfo {volume} {101}},\ \bibinfo {pages} {246806} (\bibinfo {year} {2008}{\natexlab{b}})}\BibitemShut {NoStop}%
\bibitem [{\citenamefont {Bernevig}\ and\ \citenamefont {Regnault}(2009)}]{Bernevig3}%
  \BibitemOpen
  \bibfield  {author} {\bibinfo {author} {\bibfnamefont {B.~A.}\ \bibnamefont {Bernevig}}\ and\ \bibinfo {author} {\bibfnamefont {N.}~\bibnamefont {Regnault}},\ }\bibfield  {title} {\bibinfo {title} {Anatomy of abelian and non-abelian fractional quantum hall states},\ }\href {https://doi.org/10.1103/PhysRevLett.103.206801} {\bibfield  {journal} {\bibinfo  {journal} {Phys. Rev. Lett.}\ }\textbf {\bibinfo {volume} {103}},\ \bibinfo {pages} {206801} (\bibinfo {year} {2009})}\BibitemShut {NoStop}%
\bibitem [{\citenamefont {He}\ \emph {et~al.}(2019)\citenamefont {He}, \citenamefont {Luo}, \citenamefont {Wang},\ and\ \citenamefont {Gong}}]{HeAL3}%
  \BibitemOpen
  \bibfield  {author} {\bibinfo {author} {\bibfnamefont {A.-L.}\ \bibnamefont {He}}, \bibinfo {author} {\bibfnamefont {W.-W.}\ \bibnamefont {Luo}}, \bibinfo {author} {\bibfnamefont {Y.-F.}\ \bibnamefont {Wang}},\ and\ \bibinfo {author} {\bibfnamefont {C.-D.}\ \bibnamefont {Gong}},\ }\bibfield  {title} {\bibinfo {title} {Fractional chern insulators in singular geometries},\ }\href {https://doi.org/10.1103/PhysRevB.99.165105} {\bibfield  {journal} {\bibinfo  {journal} {Phys. Rev. B}\ }\textbf {\bibinfo {volume} {99}},\ \bibinfo {pages} {165105} (\bibinfo {year} {2019})}\BibitemShut {NoStop}%
\bibitem [{\citenamefont {He}(2021)}]{HeAL4}%
  \BibitemOpen
  \bibfield  {author} {\bibinfo {author} {\bibfnamefont {A.-L.}\ \bibnamefont {He}},\ }\bibfield  {title} {\bibinfo {title} {Role of the defect-core orbital in fractional chern insulators},\ }\href {https://doi.org/10.1103/PhysRevB.103.115138} {\bibfield  {journal} {\bibinfo  {journal} {Phys. Rev. B}\ }\textbf {\bibinfo {volume} {103}},\ \bibinfo {pages} {115138} (\bibinfo {year} {2021})}\BibitemShut {NoStop}%
\bibitem [{\citenamefont {Maciejko}\ and\ \citenamefont {Rayan}(2021)}]{Joseph}%
  \BibitemOpen
  \bibfield  {author} {\bibinfo {author} {\bibfnamefont {J.}~\bibnamefont {Maciejko}}\ and\ \bibinfo {author} {\bibfnamefont {S.}~\bibnamefont {Rayan}},\ }\bibfield  {title} {\bibinfo {title} {Hyperbolic band theory},\ }\href {https://doi.org/10.1126/sciadv.abe9170} {\bibfield  {journal} {\bibinfo  {journal} {Science Advances}\ }\textbf {\bibinfo {volume} {7}},\ \bibinfo {pages} {eabe9170} (\bibinfo {year} {2021})},\ \Eprint {https://arxiv.org/abs/https://www.science.org/doi/pdf/10.1126/sciadv.abe9170} {https://www.science.org/doi/pdf/10.1126/sciadv.abe9170} \BibitemShut {NoStop}%
\bibitem [{\citenamefont {Maciejko}\ and\ \citenamefont {Rayan}(2022)}]{Joseph1}%
  \BibitemOpen
  \bibfield  {author} {\bibinfo {author} {\bibfnamefont {J.}~\bibnamefont {Maciejko}}\ and\ \bibinfo {author} {\bibfnamefont {S.}~\bibnamefont {Rayan}},\ }\bibfield  {title} {\bibinfo {title} {Automorphic bloch theorems for hyperbolic lattices},\ }\href {https://doi.org/10.1073/pnas.2116869119} {\bibfield  {journal} {\bibinfo  {journal} {Proceedings of the National Academy of Sciences}\ }\textbf {\bibinfo {volume} {119}},\ \bibinfo {pages} {e2116869119} (\bibinfo {year} {2022})},\ \Eprint {https://arxiv.org/abs/https://www.pnas.org/doi/pdf/10.1073/pnas.2116869119} {https://www.pnas.org/doi/pdf/10.1073/pnas.2116869119} \BibitemShut {NoStop}%
\bibitem [{\citenamefont {Boettcher}\ \emph {et~al.}(2022)\citenamefont {Boettcher}, \citenamefont {Gorshkov}, \citenamefont {Koll\'ar}, \citenamefont {Maciejko}, \citenamefont {Rayan},\ and\ \citenamefont {Thomale}}]{Boettcher}%
  \BibitemOpen
  \bibfield  {author} {\bibinfo {author} {\bibfnamefont {I.}~\bibnamefont {Boettcher}}, \bibinfo {author} {\bibfnamefont {A.~V.}\ \bibnamefont {Gorshkov}}, \bibinfo {author} {\bibfnamefont {A.~J.}\ \bibnamefont {Koll\'ar}}, \bibinfo {author} {\bibfnamefont {J.}~\bibnamefont {Maciejko}}, \bibinfo {author} {\bibfnamefont {S.}~\bibnamefont {Rayan}},\ and\ \bibinfo {author} {\bibfnamefont {R.}~\bibnamefont {Thomale}},\ }\bibfield  {title} {\bibinfo {title} {Crystallography of hyperbolic lattices},\ }\href {https://doi.org/10.1103/PhysRevB.105.125118} {\bibfield  {journal} {\bibinfo  {journal} {Phys. Rev. B}\ }\textbf {\bibinfo {volume} {105}},\ \bibinfo {pages} {125118} (\bibinfo {year} {2022})}\BibitemShut {NoStop}%
\bibitem [{\citenamefont {Cheng}\ \emph {et~al.}(2022)\citenamefont {Cheng}, \citenamefont {Serafin}, \citenamefont {McInerney}, \citenamefont {Rocklin}, \citenamefont {Sun},\ and\ \citenamefont {Mao}}]{ChengN}%
  \BibitemOpen
  \bibfield  {author} {\bibinfo {author} {\bibfnamefont {N.}~\bibnamefont {Cheng}}, \bibinfo {author} {\bibfnamefont {F.}~\bibnamefont {Serafin}}, \bibinfo {author} {\bibfnamefont {J.}~\bibnamefont {McInerney}}, \bibinfo {author} {\bibfnamefont {Z.}~\bibnamefont {Rocklin}}, \bibinfo {author} {\bibfnamefont {K.}~\bibnamefont {Sun}},\ and\ \bibinfo {author} {\bibfnamefont {X.}~\bibnamefont {Mao}},\ }\bibfield  {title} {\bibinfo {title} {Band theory and boundary modes of high-dimensional representations of infinite hyperbolic lattices},\ }\href {https://doi.org/10.1103/PhysRevLett.129.088002} {\bibfield  {journal} {\bibinfo  {journal} {Phys. Rev. Lett.}\ }\textbf {\bibinfo {volume} {129}},\ \bibinfo {pages} {088002} (\bibinfo {year} {2022})}\BibitemShut {NoStop}%
\bibitem [{\citenamefont {Ikeda}\ \emph {et~al.}(2023)\citenamefont {Ikeda}, \citenamefont {Matsuki},\ and\ \citenamefont {Aoki}}]{Kazuki}%
  \BibitemOpen
  \bibfield  {author} {\bibinfo {author} {\bibfnamefont {K.}~\bibnamefont {Ikeda}}, \bibinfo {author} {\bibfnamefont {Y.}~\bibnamefont {Matsuki}},\ and\ \bibinfo {author} {\bibfnamefont {S.}~\bibnamefont {Aoki}},\ }\bibfield  {title} {\bibinfo {title} {Algebra of hyperbolic band theory under magnetic field},\ }\href {https://doi.org/10.1139/cjp-2022-0145} {\bibfield  {journal} {\bibinfo  {journal} {Canadian Journal of Physics}\ }\textbf {\bibinfo {volume} {101}},\ \bibinfo {pages} {630} (\bibinfo {year} {2023})},\ \Eprint {https://arxiv.org/abs/https://doi.org/10.1139/cjp-2022-0145} {https://doi.org/10.1139/cjp-2022-0145} \BibitemShut {NoStop}%
\bibitem [{\citenamefont {Kienzle}\ and\ \citenamefont {Rayan}(2022)}]{Elliot}%
  \BibitemOpen
  \bibfield  {author} {\bibinfo {author} {\bibfnamefont {E.}~\bibnamefont {Kienzle}}\ and\ \bibinfo {author} {\bibfnamefont {S.}~\bibnamefont {Rayan}},\ }\bibfield  {title} {\bibinfo {title} {Hyperbolic band theory through higgs bundles},\ }\href {https://doi.org/https://doi.org/10.1016/j.aim.2022.108664} {\bibfield  {journal} {\bibinfo  {journal} {Advances in Mathematics}\ }\textbf {\bibinfo {volume} {409}},\ \bibinfo {pages} {108664} (\bibinfo {year} {2022})}\BibitemShut {NoStop}%
\bibitem [{\citenamefont {Lenggenhager}\ \emph {et~al.}(2023)\citenamefont {Lenggenhager}, \citenamefont {Maciejko},\ and\ \citenamefont {Bzdu\ifmmode~\check{s}\else \v{s}\fi{}ek}}]{Patrick}%
  \BibitemOpen
  \bibfield  {author} {\bibinfo {author} {\bibfnamefont {P.~M.}\ \bibnamefont {Lenggenhager}}, \bibinfo {author} {\bibfnamefont {J.}~\bibnamefont {Maciejko}},\ and\ \bibinfo {author} {\bibfnamefont {T.~c.~v.}\ \bibnamefont {Bzdu\ifmmode~\check{s}\else \v{s}\fi{}ek}},\ }\bibfield  {title} {\bibinfo {title} {Non-abelian hyperbolic band theory from supercells},\ }\href {https://doi.org/10.1103/PhysRevLett.131.226401} {\bibfield  {journal} {\bibinfo  {journal} {Phys. Rev. Lett.}\ }\textbf {\bibinfo {volume} {131}},\ \bibinfo {pages} {226401} (\bibinfo {year} {2023})}\BibitemShut {NoStop}%
\bibitem [{\citenamefont {Lux}\ and\ \citenamefont {Prodan}(2023)}]{Lux}%
  \BibitemOpen
  \bibfield  {author} {\bibinfo {author} {\bibfnamefont {F.~R.}\ \bibnamefont {Lux}}\ and\ \bibinfo {author} {\bibfnamefont {E.}~\bibnamefont {Prodan}},\ }\bibfield  {title} {\bibinfo {title} {Converging periodic boundary conditions and detection of topological gaps on regular hyperbolic tessellations},\ }\href {https://doi.org/10.1103/PhysRevLett.131.176603} {\bibfield  {journal} {\bibinfo  {journal} {Phys. Rev. Lett.}\ }\textbf {\bibinfo {volume} {131}},\ \bibinfo {pages} {176603} (\bibinfo {year} {2023})}\BibitemShut {NoStop}%
\bibitem [{\citenamefont {Mosseri}\ and\ \citenamefont {Vidal}(2023)}]{Mosseri}%
  \BibitemOpen
  \bibfield  {author} {\bibinfo {author} {\bibfnamefont {R.}~\bibnamefont {Mosseri}}\ and\ \bibinfo {author} {\bibfnamefont {J.}~\bibnamefont {Vidal}},\ }\bibfield  {title} {\bibinfo {title} {Density of states of tight-binding models in the hyperbolic plane},\ }\href {https://doi.org/10.1103/PhysRevB.108.035154} {\bibfield  {journal} {\bibinfo  {journal} {Phys. Rev. B}\ }\textbf {\bibinfo {volume} {108}},\ \bibinfo {pages} {035154} (\bibinfo {year} {2023})}\BibitemShut {NoStop}%
\bibitem [{\citenamefont {Gluscevich}\ \emph {et~al.}(2023)\citenamefont {Gluscevich}, \citenamefont {Samanta}, \citenamefont {Manna},\ and\ \citenamefont {Roy}}]{NobleG}%
  \BibitemOpen
  \bibfield  {author} {\bibinfo {author} {\bibfnamefont {N.}~\bibnamefont {Gluscevich}}, \bibinfo {author} {\bibfnamefont {A.}~\bibnamefont {Samanta}}, \bibinfo {author} {\bibfnamefont {S.}~\bibnamefont {Manna}},\ and\ \bibinfo {author} {\bibfnamefont {B.}~\bibnamefont {Roy}},\ }\href@noop {} {\bibinfo {title} {Dynamic mass generation on two-dimensional electronic hyperbolic lattices}} (\bibinfo {year} {2023}),\ \Eprint {https://arxiv.org/abs/2302.04864} {arXiv:2302.04864 [cond-mat.str-el]} \BibitemShut {NoStop}%
\bibitem [{\citenamefont {Bzdu\ifmmode~\check{s}\else \v{s}\fi{}ek}\ and\ \citenamefont {Maciejko}(2022)}]{FBH1}%
  \BibitemOpen
  \bibfield  {author} {\bibinfo {author} {\bibfnamefont {T.~c.~v.}\ \bibnamefont {Bzdu\ifmmode~\check{s}\else \v{s}\fi{}ek}}\ and\ \bibinfo {author} {\bibfnamefont {J.}~\bibnamefont {Maciejko}},\ }\bibfield  {title} {\bibinfo {title} {Flat bands and band-touching from real-space topology in hyperbolic lattices},\ }\href {https://doi.org/10.1103/PhysRevB.106.155146} {\bibfield  {journal} {\bibinfo  {journal} {Phys. Rev. B}\ }\textbf {\bibinfo {volume} {106}},\ \bibinfo {pages} {155146} (\bibinfo {year} {2022})}\BibitemShut {NoStop}%
\bibitem [{\citenamefont {Mosseri}\ \emph {et~al.}(2022)\citenamefont {Mosseri}, \citenamefont {Vogeler},\ and\ \citenamefont {Vidal}}]{FBH2}%
  \BibitemOpen
  \bibfield  {author} {\bibinfo {author} {\bibfnamefont {R.}~\bibnamefont {Mosseri}}, \bibinfo {author} {\bibfnamefont {R.}~\bibnamefont {Vogeler}},\ and\ \bibinfo {author} {\bibfnamefont {J.}~\bibnamefont {Vidal}},\ }\bibfield  {title} {\bibinfo {title} {Aharonov-bohm cages, flat bands, and gap labeling in hyperbolic tilings},\ }\href {https://doi.org/10.1103/PhysRevB.106.155120} {\bibfield  {journal} {\bibinfo  {journal} {Phys. Rev. B}\ }\textbf {\bibinfo {volume} {106}},\ \bibinfo {pages} {155120} (\bibinfo {year} {2022})}\BibitemShut {NoStop}%
\bibitem [{\citenamefont {Yuan}\ \emph {et~al.}(2024)\citenamefont {Yuan}, \citenamefont {Zhang}, \citenamefont {Pei},\ and\ \citenamefont {Zhang}}]{TFBZhang}%
  \BibitemOpen
  \bibfield  {author} {\bibinfo {author} {\bibfnamefont {H.}~\bibnamefont {Yuan}}, \bibinfo {author} {\bibfnamefont {W.}~\bibnamefont {Zhang}}, \bibinfo {author} {\bibfnamefont {Q.}~\bibnamefont {Pei}},\ and\ \bibinfo {author} {\bibfnamefont {X.}~\bibnamefont {Zhang}},\ }\bibfield  {title} {\bibinfo {title} {Hyperbolic topological flat bands},\ }\href {https://doi.org/10.1103/PhysRevB.109.L041109} {\bibfield  {journal} {\bibinfo  {journal} {Phys. Rev. B}\ }\textbf {\bibinfo {volume} {109}},\ \bibinfo {pages} {L041109} (\bibinfo {year} {2024})}\BibitemShut {NoStop}%
\bibitem [{\citenamefont {Guan}\ \emph {et~al.}(2024)\citenamefont {Guan}, \citenamefont {Qi}, \citenamefont {Zhou}, \citenamefont {He},\ and\ \citenamefont {Wang}}]{Guanpre}%
  \BibitemOpen
  \bibfield  {author} {\bibinfo {author} {\bibfnamefont {D.-H.}\ \bibnamefont {Guan}}, \bibinfo {author} {\bibfnamefont {L.}~\bibnamefont {Qi}}, \bibinfo {author} {\bibfnamefont {Y.}~\bibnamefont {Zhou}}, \bibinfo {author} {\bibfnamefont {A.-L.}\ \bibnamefont {He}},\ and\ \bibinfo {author} {\bibfnamefont {Y.-F.}\ \bibnamefont {Wang}},\ }\href {https://arxiv.org/abs/2408.16615} {\bibinfo {title} {Topological flat bands in hyperbolic lattices}} (\bibinfo {year} {2024}),\ \Eprint {https://arxiv.org/abs/2408.16615} {arXiv:2408.16615 [cond-mat.str-el]} \BibitemShut {NoStop}%
\bibitem [{\citenamefont {Curtis}\ \emph {et~al.}(2023)\citenamefont {Curtis}, \citenamefont {Narang},\ and\ \citenamefont {Galitski}}]{curtis2023}%
  \BibitemOpen
  \bibfield  {author} {\bibinfo {author} {\bibfnamefont {J.~B.}\ \bibnamefont {Curtis}}, \bibinfo {author} {\bibfnamefont {P.}~\bibnamefont {Narang}},\ and\ \bibinfo {author} {\bibfnamefont {V.}~\bibnamefont {Galitski}},\ }\href@noop {} {\bibinfo {title} {Absence of weak localization on negative curvature surfaces}} (\bibinfo {year} {2023}),\ \Eprint {https://arxiv.org/abs/2308.01351} {arXiv:2308.01351 [cond-mat.dis-nn]} \BibitemShut {NoStop}%
\bibitem [{\citenamefont {Chen}\ \emph {et~al.}(2023{\natexlab{a}})\citenamefont {Chen}, \citenamefont {Maciejko},\ and\ \citenamefont {Boettcher}}]{chenA}%
  \BibitemOpen
  \bibfield  {author} {\bibinfo {author} {\bibfnamefont {A.}~\bibnamefont {Chen}}, \bibinfo {author} {\bibfnamefont {J.}~\bibnamefont {Maciejko}},\ and\ \bibinfo {author} {\bibfnamefont {I.}~\bibnamefont {Boettcher}},\ }\href@noop {} {\bibinfo {title} {Anderson localization transition in disordered hyperbolic lattices}} (\bibinfo {year} {2023}{\natexlab{a}}),\ \Eprint {https://arxiv.org/abs/2310.07978} {arXiv:2310.07978 [cond-mat.dis-nn]} \BibitemShut {NoStop}%
\bibitem [{\citenamefont {Zhu}\ \emph {et~al.}(2021)\citenamefont {Zhu}, \citenamefont {Guo}, \citenamefont {Breuckmann}, \citenamefont {Guo},\ and\ \citenamefont {Feng}}]{Zhuxc}%
  \BibitemOpen
  \bibfield  {author} {\bibinfo {author} {\bibfnamefont {X.}~\bibnamefont {Zhu}}, \bibinfo {author} {\bibfnamefont {J.}~\bibnamefont {Guo}}, \bibinfo {author} {\bibfnamefont {N.~P.}\ \bibnamefont {Breuckmann}}, \bibinfo {author} {\bibfnamefont {H.}~\bibnamefont {Guo}},\ and\ \bibinfo {author} {\bibfnamefont {S.}~\bibnamefont {Feng}},\ }\bibfield  {title} {\bibinfo {title} {Quantum phase transitions of interacting bosons on hyperbolic lattices},\ }\href {https://doi.org/10.1088/1361-648X/ac0a1a} {\bibfield  {journal} {\bibinfo  {journal} {Journal of Physics: Condensed Matter}\ }\textbf {\bibinfo {volume} {33}},\ \bibinfo {pages} {335602} (\bibinfo {year} {2021})}\BibitemShut {NoStop}%
\bibitem [{\citenamefont {Bienias}\ \emph {et~al.}(2022)\citenamefont {Bienias}, \citenamefont {Boettcher}, \citenamefont {Belyansky}, \citenamefont {Koll\'ar},\ and\ \citenamefont {Gorshkov}}]{BieniasP}%
  \BibitemOpen
  \bibfield  {author} {\bibinfo {author} {\bibfnamefont {P.}~\bibnamefont {Bienias}}, \bibinfo {author} {\bibfnamefont {I.}~\bibnamefont {Boettcher}}, \bibinfo {author} {\bibfnamefont {R.}~\bibnamefont {Belyansky}}, \bibinfo {author} {\bibfnamefont {A.~J.}\ \bibnamefont {Koll\'ar}},\ and\ \bibinfo {author} {\bibfnamefont {A.~V.}\ \bibnamefont {Gorshkov}},\ }\bibfield  {title} {\bibinfo {title} {Circuit quantum electrodynamics in hyperbolic space: From photon bound states to frustrated spin models},\ }\href {https://doi.org/10.1103/PhysRevLett.128.013601} {\bibfield  {journal} {\bibinfo  {journal} {Phys. Rev. Lett.}\ }\textbf {\bibinfo {volume} {128}},\ \bibinfo {pages} {013601} (\bibinfo {year} {2022})}\BibitemShut {NoStop}%
\bibitem [{\citenamefont {Gluscevich}\ and\ \citenamefont {Roy}(2023)}]{NobleG2}%
  \BibitemOpen
  \bibfield  {author} {\bibinfo {author} {\bibfnamefont {N.}~\bibnamefont {Gluscevich}}\ and\ \bibinfo {author} {\bibfnamefont {B.}~\bibnamefont {Roy}},\ }\href@noop {} {\bibinfo {title} {Magnetic catalysis in weakly interacting hyperbolic dirac materials}} (\bibinfo {year} {2023}),\ \Eprint {https://arxiv.org/abs/2305.11174} {arXiv:2305.11174 [cond-mat.str-el]} \BibitemShut {NoStop}%
\bibitem [{\citenamefont {Ikeda}\ \emph {et~al.}(2021)\citenamefont {Ikeda}, \citenamefont {Aoki},\ and\ \citenamefont {Matsuki}}]{Kazuki2}%
  \BibitemOpen
  \bibfield  {author} {\bibinfo {author} {\bibfnamefont {K.}~\bibnamefont {Ikeda}}, \bibinfo {author} {\bibfnamefont {S.}~\bibnamefont {Aoki}},\ and\ \bibinfo {author} {\bibfnamefont {Y.}~\bibnamefont {Matsuki}},\ }\bibfield  {title} {\bibinfo {title} {Hyperbolic band theory under magnetic field and dirac cones on a higher genus surface},\ }\href {https://doi.org/10.1088/1361-648X/ac24c4} {\bibfield  {journal} {\bibinfo  {journal} {Journal of Physics: Condensed Matter}\ }\textbf {\bibinfo {volume} {33}},\ \bibinfo {pages} {485602} (\bibinfo {year} {2021})}\BibitemShut {NoStop}%
\bibitem [{\citenamefont {Yu}\ \emph {et~al.}(2020)\citenamefont {Yu}, \citenamefont {Piao},\ and\ \citenamefont {Park}}]{YuS}%
  \BibitemOpen
  \bibfield  {author} {\bibinfo {author} {\bibfnamefont {S.}~\bibnamefont {Yu}}, \bibinfo {author} {\bibfnamefont {X.}~\bibnamefont {Piao}},\ and\ \bibinfo {author} {\bibfnamefont {N.}~\bibnamefont {Park}},\ }\bibfield  {title} {\bibinfo {title} {Topological hyperbolic lattices},\ }\href {https://doi.org/10.1103/PhysRevLett.125.053901} {\bibfield  {journal} {\bibinfo  {journal} {Phys. Rev. Lett.}\ }\textbf {\bibinfo {volume} {125}},\ \bibinfo {pages} {053901} (\bibinfo {year} {2020})}\BibitemShut {NoStop}%
\bibitem [{\citenamefont {Stegmaier}\ \emph {et~al.}(2022)\citenamefont {Stegmaier}, \citenamefont {Upreti}, \citenamefont {Thomale},\ and\ \citenamefont {Boettcher}}]{Stegmaier}%
  \BibitemOpen
  \bibfield  {author} {\bibinfo {author} {\bibfnamefont {A.}~\bibnamefont {Stegmaier}}, \bibinfo {author} {\bibfnamefont {L.~K.}\ \bibnamefont {Upreti}}, \bibinfo {author} {\bibfnamefont {R.}~\bibnamefont {Thomale}},\ and\ \bibinfo {author} {\bibfnamefont {I.}~\bibnamefont {Boettcher}},\ }\bibfield  {title} {\bibinfo {title} {Universality of hofstadter butterflies on hyperbolic lattices},\ }\href {https://doi.org/10.1103/PhysRevLett.128.166402} {\bibfield  {journal} {\bibinfo  {journal} {Phys. Rev. Lett.}\ }\textbf {\bibinfo {volume} {128}},\ \bibinfo {pages} {166402} (\bibinfo {year} {2022})}\BibitemShut {NoStop}%
\bibitem [{\citenamefont {Urwyler}\ \emph {et~al.}(2022)\citenamefont {Urwyler}, \citenamefont {Lenggenhager}, \citenamefont {Boettcher}, \citenamefont {Thomale}, \citenamefont {Neupert},\ and\ \citenamefont {Bzdu\ifmmode~\check{s}\else \v{s}\fi{}ek}}]{HyperTI11}%
  \BibitemOpen
  \bibfield  {author} {\bibinfo {author} {\bibfnamefont {D.~M.}\ \bibnamefont {Urwyler}}, \bibinfo {author} {\bibfnamefont {P.~M.}\ \bibnamefont {Lenggenhager}}, \bibinfo {author} {\bibfnamefont {I.}~\bibnamefont {Boettcher}}, \bibinfo {author} {\bibfnamefont {R.}~\bibnamefont {Thomale}}, \bibinfo {author} {\bibfnamefont {T.}~\bibnamefont {Neupert}},\ and\ \bibinfo {author} {\bibfnamefont {T.~c.~v.}\ \bibnamefont {Bzdu\ifmmode~\check{s}\else \v{s}\fi{}ek}},\ }\bibfield  {title} {\bibinfo {title} {Hyperbolic topological band insulators},\ }\href {https://doi.org/10.1103/PhysRevLett.129.246402} {\bibfield  {journal} {\bibinfo  {journal} {Phys. Rev. Lett.}\ }\textbf {\bibinfo {volume} {129}},\ \bibinfo {pages} {246402} (\bibinfo {year} {2022})}\BibitemShut {NoStop}%
\bibitem [{\citenamefont {Zhang}\ \emph {et~al.}(2022)\citenamefont {Zhang}, \citenamefont {Yuan}, \citenamefont {Sun}, \citenamefont {Sun},\ and\ \citenamefont {Zhang}}]{Zhang2022}%
  \BibitemOpen
  \bibfield  {author} {\bibinfo {author} {\bibfnamefont {W.}~\bibnamefont {Zhang}}, \bibinfo {author} {\bibfnamefont {H.}~\bibnamefont {Yuan}}, \bibinfo {author} {\bibfnamefont {N.}~\bibnamefont {Sun}}, \bibinfo {author} {\bibfnamefont {H.}~\bibnamefont {Sun}},\ and\ \bibinfo {author} {\bibfnamefont {X.}~\bibnamefont {Zhang}},\ }\bibfield  {title} {\bibinfo {title} {Observation of novel topological states in hyperbolic lattices},\ }\href {https://doi.org/10.1038/s41467-022-30631-x} {\bibfield  {journal} {\bibinfo  {journal} {Nature Communications}\ }\textbf {\bibinfo {volume} {13}},\ \bibinfo {pages} {2937} (\bibinfo {year} {2022})}\BibitemShut {NoStop}%
\bibitem [{\citenamefont {Liu}\ \emph {et~al.}(2022)\citenamefont {Liu}, \citenamefont {Hua}, \citenamefont {Peng},\ and\ \citenamefont {Zhou}}]{LiuZR}%
  \BibitemOpen
  \bibfield  {author} {\bibinfo {author} {\bibfnamefont {Z.-R.}\ \bibnamefont {Liu}}, \bibinfo {author} {\bibfnamefont {C.-B.}\ \bibnamefont {Hua}}, \bibinfo {author} {\bibfnamefont {T.}~\bibnamefont {Peng}},\ and\ \bibinfo {author} {\bibfnamefont {B.}~\bibnamefont {Zhou}},\ }\bibfield  {title} {\bibinfo {title} {Chern insulator in a hyperbolic lattice},\ }\href {https://doi.org/10.1103/PhysRevB.105.245301} {\bibfield  {journal} {\bibinfo  {journal} {Phys. Rev. B}\ }\textbf {\bibinfo {volume} {105}},\ \bibinfo {pages} {245301} (\bibinfo {year} {2022})}\BibitemShut {NoStop}%
\bibitem [{\citenamefont {Zhang}\ \emph {et~al.}(2023)\citenamefont {Zhang}, \citenamefont {Di}, \citenamefont {Zheng}, \citenamefont {Sun},\ and\ \citenamefont {Zhang}}]{Zhang2023}%
  \BibitemOpen
  \bibfield  {author} {\bibinfo {author} {\bibfnamefont {W.}~\bibnamefont {Zhang}}, \bibinfo {author} {\bibfnamefont {F.}~\bibnamefont {Di}}, \bibinfo {author} {\bibfnamefont {X.}~\bibnamefont {Zheng}}, \bibinfo {author} {\bibfnamefont {H.}~\bibnamefont {Sun}},\ and\ \bibinfo {author} {\bibfnamefont {X.}~\bibnamefont {Zhang}},\ }\bibfield  {title} {\bibinfo {title} {Hyperbolic band topology with non-trivial second chern numbers},\ }\href {https://doi.org/10.1038/s41467-023-36767-8} {\bibfield  {journal} {\bibinfo  {journal} {Nature Communications}\ }\textbf {\bibinfo {volume} {14}},\ \bibinfo {pages} {1083} (\bibinfo {year} {2023})}\BibitemShut {NoStop}%
\bibitem [{\citenamefont {Liu}\ \emph {et~al.}(2023)\citenamefont {Liu}, \citenamefont {Hua}, \citenamefont {Peng}, \citenamefont {Chen},\ and\ \citenamefont {Zhou}}]{LiuZR1}%
  \BibitemOpen
  \bibfield  {author} {\bibinfo {author} {\bibfnamefont {Z.-R.}\ \bibnamefont {Liu}}, \bibinfo {author} {\bibfnamefont {C.-B.}\ \bibnamefont {Hua}}, \bibinfo {author} {\bibfnamefont {T.}~\bibnamefont {Peng}}, \bibinfo {author} {\bibfnamefont {R.}~\bibnamefont {Chen}},\ and\ \bibinfo {author} {\bibfnamefont {B.}~\bibnamefont {Zhou}},\ }\bibfield  {title} {\bibinfo {title} {Higher-order topological insulators in hyperbolic lattices},\ }\href {https://doi.org/10.1103/PhysRevB.107.125302} {\bibfield  {journal} {\bibinfo  {journal} {Phys. Rev. B}\ }\textbf {\bibinfo {volume} {107}},\ \bibinfo {pages} {125302} (\bibinfo {year} {2023})}\BibitemShut {NoStop}%
\bibitem [{\citenamefont {Pei}\ \emph {et~al.}(2023)\citenamefont {Pei}, \citenamefont {Yuan}, \citenamefont {Zhang},\ and\ \citenamefont {Zhang}}]{PeiQS}%
  \BibitemOpen
  \bibfield  {author} {\bibinfo {author} {\bibfnamefont {Q.}~\bibnamefont {Pei}}, \bibinfo {author} {\bibfnamefont {H.}~\bibnamefont {Yuan}}, \bibinfo {author} {\bibfnamefont {W.}~\bibnamefont {Zhang}},\ and\ \bibinfo {author} {\bibfnamefont {X.}~\bibnamefont {Zhang}},\ }\bibfield  {title} {\bibinfo {title} {Engineering boundary-dominated topological states in defective hyperbolic lattices},\ }\href {https://doi.org/10.1103/PhysRevB.107.165145} {\bibfield  {journal} {\bibinfo  {journal} {Phys. Rev. B}\ }\textbf {\bibinfo {volume} {107}},\ \bibinfo {pages} {165145} (\bibinfo {year} {2023})}\BibitemShut {NoStop}%
\bibitem [{\citenamefont {Tao}\ and\ \citenamefont {Xu}(2023)}]{TaoYL}%
  \BibitemOpen
  \bibfield  {author} {\bibinfo {author} {\bibfnamefont {Y.-L.}\ \bibnamefont {Tao}}\ and\ \bibinfo {author} {\bibfnamefont {Y.}~\bibnamefont {Xu}},\ }\bibfield  {title} {\bibinfo {title} {Higher-order topological hyperbolic lattices},\ }\href {https://doi.org/10.1103/PhysRevB.107.184201} {\bibfield  {journal} {\bibinfo  {journal} {Phys. Rev. B}\ }\textbf {\bibinfo {volume} {107}},\ \bibinfo {pages} {184201} (\bibinfo {year} {2023})}\BibitemShut {NoStop}%
\bibitem [{\citenamefont {Tummuru}\ \emph {et~al.}(2024)\citenamefont {Tummuru}, \citenamefont {Chen}, \citenamefont {Lenggenhager}, \citenamefont {Neupert}, \citenamefont {Maciejko},\ and\ \citenamefont {Bzdu\ifmmode~\check{s}\else \v{s}\fi{}ek}}]{TarunT}%
  \BibitemOpen
  \bibfield  {author} {\bibinfo {author} {\bibfnamefont {T.}~\bibnamefont {Tummuru}}, \bibinfo {author} {\bibfnamefont {A.}~\bibnamefont {Chen}}, \bibinfo {author} {\bibfnamefont {P.~M.}\ \bibnamefont {Lenggenhager}}, \bibinfo {author} {\bibfnamefont {T.}~\bibnamefont {Neupert}}, \bibinfo {author} {\bibfnamefont {J.}~\bibnamefont {Maciejko}},\ and\ \bibinfo {author} {\bibfnamefont {T.~c.~v.}\ \bibnamefont {Bzdu\ifmmode~\check{s}\else \v{s}\fi{}ek}},\ }\bibfield  {title} {\bibinfo {title} {Hyperbolic non-abelian semimetal},\ }\href {https://doi.org/10.1103/PhysRevLett.132.206601} {\bibfield  {journal} {\bibinfo  {journal} {Phys. Rev. Lett.}\ }\textbf {\bibinfo {volume} {132}},\ \bibinfo {pages} {206601} (\bibinfo {year} {2024})}\BibitemShut {NoStop}%
\bibitem [{\citenamefont {Chen}\ \emph {et~al.}(2023{\natexlab{b}})\citenamefont {Chen}, \citenamefont {Guan}, \citenamefont {Lenggenhager}, \citenamefont {Maciejko}, \citenamefont {Boettcher},\ and\ \citenamefont {Bzdu\ifmmode~\check{s}\else \v{s}\fi{}ek}}]{Anffany}%
  \BibitemOpen
  \bibfield  {author} {\bibinfo {author} {\bibfnamefont {A.}~\bibnamefont {Chen}}, \bibinfo {author} {\bibfnamefont {Y.}~\bibnamefont {Guan}}, \bibinfo {author} {\bibfnamefont {P.~M.}\ \bibnamefont {Lenggenhager}}, \bibinfo {author} {\bibfnamefont {J.}~\bibnamefont {Maciejko}}, \bibinfo {author} {\bibfnamefont {I.}~\bibnamefont {Boettcher}},\ and\ \bibinfo {author} {\bibfnamefont {T.~c.~v.}\ \bibnamefont {Bzdu\ifmmode~\check{s}\else \v{s}\fi{}ek}},\ }\bibfield  {title} {\bibinfo {title} {Symmetry and topology of hyperbolic haldane models},\ }\href {https://doi.org/10.1103/PhysRevB.108.085114} {\bibfield  {journal} {\bibinfo  {journal} {Phys. Rev. B}\ }\textbf {\bibinfo {volume} {108}},\ \bibinfo {pages} {085114} (\bibinfo {year} {2023}{\natexlab{b}})}\BibitemShut {NoStop}%
\bibitem [{\citenamefont {Sun}\ \emph {et~al.}(2023)\citenamefont {Sun}, \citenamefont {Li}, \citenamefont {Feng},\ and\ \citenamefont {Guo}}]{SunJ}%
  \BibitemOpen
  \bibfield  {author} {\bibinfo {author} {\bibfnamefont {J.}~\bibnamefont {Sun}}, \bibinfo {author} {\bibfnamefont {C.-A.}\ \bibnamefont {Li}}, \bibinfo {author} {\bibfnamefont {S.}~\bibnamefont {Feng}},\ and\ \bibinfo {author} {\bibfnamefont {H.}~\bibnamefont {Guo}},\ }\bibfield  {title} {\bibinfo {title} {Hybrid higher-order skin-topological effect in hyperbolic lattices},\ }\href {https://doi.org/10.1103/PhysRevB.108.075122} {\bibfield  {journal} {\bibinfo  {journal} {Phys. Rev. B}\ }\textbf {\bibinfo {volume} {108}},\ \bibinfo {pages} {075122} (\bibinfo {year} {2023})}\BibitemShut {NoStop}%
\bibitem [{\citenamefont {Basteiro}\ \emph {et~al.}(2023)\citenamefont {Basteiro}, \citenamefont {Dusel}, \citenamefont {Erdmenger}, \citenamefont {Herdt}, \citenamefont {Hinrichsen}, \citenamefont {Meyer},\ and\ \citenamefont {Schrauth}}]{Basteiro}%
  \BibitemOpen
  \bibfield  {author} {\bibinfo {author} {\bibfnamefont {P.}~\bibnamefont {Basteiro}}, \bibinfo {author} {\bibfnamefont {F.}~\bibnamefont {Dusel}}, \bibinfo {author} {\bibfnamefont {J.}~\bibnamefont {Erdmenger}}, \bibinfo {author} {\bibfnamefont {D.}~\bibnamefont {Herdt}}, \bibinfo {author} {\bibfnamefont {H.}~\bibnamefont {Hinrichsen}}, \bibinfo {author} {\bibfnamefont {R.}~\bibnamefont {Meyer}},\ and\ \bibinfo {author} {\bibfnamefont {M.}~\bibnamefont {Schrauth}},\ }\bibfield  {title} {\bibinfo {title} {Breitenlohner-freedman bound on hyperbolic tilings},\ }\href {https://doi.org/10.1103/PhysRevLett.130.091604} {\bibfield  {journal} {\bibinfo  {journal} {Phys. Rev. Lett.}\ }\textbf {\bibinfo {volume} {130}},\ \bibinfo {pages} {091604} (\bibinfo {year} {2023})}\BibitemShut {NoStop}%
\bibitem [{\citenamefont {Koll{\'a}r}\ \emph {et~al.}(2019)\citenamefont {Koll{\'a}r}, \citenamefont {Fitzpatrick},\ and\ \citenamefont {Houck}}]{Kollar2019}%
  \BibitemOpen
  \bibfield  {author} {\bibinfo {author} {\bibfnamefont {A.~J.}\ \bibnamefont {Koll{\'a}r}}, \bibinfo {author} {\bibfnamefont {M.}~\bibnamefont {Fitzpatrick}},\ and\ \bibinfo {author} {\bibfnamefont {A.~A.}\ \bibnamefont {Houck}},\ }\bibfield  {title} {\bibinfo {title} {Hyperbolic lattices in circuit quantum electrodynamics},\ }\href {https://doi.org/10.1038/s41586-019-1348-3} {\bibfield  {journal} {\bibinfo  {journal} {Nature}\ }\textbf {\bibinfo {volume} {571}},\ \bibinfo {pages} {45} (\bibinfo {year} {2019})}\BibitemShut {NoStop}%
\bibitem [{\citenamefont {Boettcher}\ \emph {et~al.}(2020)\citenamefont {Boettcher}, \citenamefont {Bienias}, \citenamefont {Belyansky}, \citenamefont {Koll\'ar},\ and\ \citenamefont {Gorshkov}}]{Boettcher1}%
  \BibitemOpen
  \bibfield  {author} {\bibinfo {author} {\bibfnamefont {I.}~\bibnamefont {Boettcher}}, \bibinfo {author} {\bibfnamefont {P.}~\bibnamefont {Bienias}}, \bibinfo {author} {\bibfnamefont {R.}~\bibnamefont {Belyansky}}, \bibinfo {author} {\bibfnamefont {A.~J.}\ \bibnamefont {Koll\'ar}},\ and\ \bibinfo {author} {\bibfnamefont {A.~V.}\ \bibnamefont {Gorshkov}},\ }\bibfield  {title} {\bibinfo {title} {Quantum simulation of hyperbolic space with circuit quantum electrodynamics: From graphs to geometry},\ }\href {https://doi.org/10.1103/PhysRevA.102.032208} {\bibfield  {journal} {\bibinfo  {journal} {Phys. Rev. A}\ }\textbf {\bibinfo {volume} {102}},\ \bibinfo {pages} {032208} (\bibinfo {year} {2020})}\BibitemShut {NoStop}%
\bibitem [{\citenamefont {Lenggenhager}\ \emph {et~al.}(2022)\citenamefont {Lenggenhager}, \citenamefont {Stegmaier}, \citenamefont {Upreti}, \citenamefont {Hofmann}, \citenamefont {Helbig}, \citenamefont {Vollhardt}, \citenamefont {Greiter}, \citenamefont {Lee}, \citenamefont {Imhof}, \citenamefont {Brand}, \citenamefont {Kie{\ss}ling}, \citenamefont {Boettcher}, \citenamefont {Neupert}, \citenamefont {Thomale},\ and\ \citenamefont {Bzdu{\v{s}}ek}}]{Lenggenhager2022}%
  \BibitemOpen
  \bibfield  {author} {\bibinfo {author} {\bibfnamefont {P.~M.}\ \bibnamefont {Lenggenhager}}, \bibinfo {author} {\bibfnamefont {A.}~\bibnamefont {Stegmaier}}, \bibinfo {author} {\bibfnamefont {L.~K.}\ \bibnamefont {Upreti}}, \bibinfo {author} {\bibfnamefont {T.}~\bibnamefont {Hofmann}}, \bibinfo {author} {\bibfnamefont {T.}~\bibnamefont {Helbig}}, \bibinfo {author} {\bibfnamefont {A.}~\bibnamefont {Vollhardt}}, \bibinfo {author} {\bibfnamefont {M.}~\bibnamefont {Greiter}}, \bibinfo {author} {\bibfnamefont {C.~H.}\ \bibnamefont {Lee}}, \bibinfo {author} {\bibfnamefont {S.}~\bibnamefont {Imhof}}, \bibinfo {author} {\bibfnamefont {H.}~\bibnamefont {Brand}}, \bibinfo {author} {\bibfnamefont {T.}~\bibnamefont {Kie{\ss}ling}}, \bibinfo {author} {\bibfnamefont {I.}~\bibnamefont {Boettcher}}, \bibinfo {author} {\bibfnamefont {T.}~\bibnamefont {Neupert}}, \bibinfo {author} {\bibfnamefont {R.}~\bibnamefont {Thomale}},\ and\ \bibinfo {author} {\bibfnamefont {T.}~\bibnamefont {Bzdu{\v{s}}ek}},\ }\bibfield  {title}
  {\bibinfo {title} {Simulating hyperbolic space on a circuit board},\ }\href {https://doi.org/10.1038/s41467-022-32042-4} {\bibfield  {journal} {\bibinfo  {journal} {Nature Communications}\ }\textbf {\bibinfo {volume} {13}},\ \bibinfo {pages} {4373} (\bibinfo {year} {2022})}\BibitemShut {NoStop}%
\bibitem [{\citenamefont {Chen}\ \emph {et~al.}(2023{\natexlab{c}})\citenamefont {Chen}, \citenamefont {Brand}, \citenamefont {Helbig}, \citenamefont {Hofmann}, \citenamefont {Imhof}, \citenamefont {Fritzsche}, \citenamefont {Kie{\ss}ling}, \citenamefont {Stegmaier}, \citenamefont {Upreti}, \citenamefont {Neupert}, \citenamefont {Bzdu{\v{s}}ek}, \citenamefont {Greiter}, \citenamefont {Thomale},\ and\ \citenamefont {Boettcher}}]{Chen2023}%
  \BibitemOpen
  \bibfield  {author} {\bibinfo {author} {\bibfnamefont {A.}~\bibnamefont {Chen}}, \bibinfo {author} {\bibfnamefont {H.}~\bibnamefont {Brand}}, \bibinfo {author} {\bibfnamefont {T.}~\bibnamefont {Helbig}}, \bibinfo {author} {\bibfnamefont {T.}~\bibnamefont {Hofmann}}, \bibinfo {author} {\bibfnamefont {S.}~\bibnamefont {Imhof}}, \bibinfo {author} {\bibfnamefont {A.}~\bibnamefont {Fritzsche}}, \bibinfo {author} {\bibfnamefont {T.}~\bibnamefont {Kie{\ss}ling}}, \bibinfo {author} {\bibfnamefont {A.}~\bibnamefont {Stegmaier}}, \bibinfo {author} {\bibfnamefont {L.~K.}\ \bibnamefont {Upreti}}, \bibinfo {author} {\bibfnamefont {T.}~\bibnamefont {Neupert}}, \bibinfo {author} {\bibfnamefont {T.}~\bibnamefont {Bzdu{\v{s}}ek}}, \bibinfo {author} {\bibfnamefont {M.}~\bibnamefont {Greiter}}, \bibinfo {author} {\bibfnamefont {R.}~\bibnamefont {Thomale}},\ and\ \bibinfo {author} {\bibfnamefont {I.}~\bibnamefont {Boettcher}},\ }\bibfield  {title} {\bibinfo {title} {Hyperbolic matter in electrical circuits with tunable complex
  phases},\ }\href {https://doi.org/10.1038/s41467-023-36359-6} {\bibfield  {journal} {\bibinfo  {journal} {Nature Communications}\ }\textbf {\bibinfo {volume} {14}},\ \bibinfo {pages} {622} (\bibinfo {year} {2023}{\natexlab{c}})}\BibitemShut {NoStop}%
\bibitem [{\citenamefont {Kj\"all}\ and\ \citenamefont {Moore}(2012)}]{JEMoore}%
  \BibitemOpen
  \bibfield  {author} {\bibinfo {author} {\bibfnamefont {J.~A.}\ \bibnamefont {Kj\"all}}\ and\ \bibinfo {author} {\bibfnamefont {J.~E.}\ \bibnamefont {Moore}},\ }\bibfield  {title} {\bibinfo {title} {Edge excitations of bosonic fractional quantum hall phases in optical lattices},\ }\href {https://doi.org/10.1103/PhysRevB.85.235137} {\bibfield  {journal} {\bibinfo  {journal} {Phys. Rev. B}\ }\textbf {\bibinfo {volume} {85}},\ \bibinfo {pages} {235137} (\bibinfo {year} {2012})}\BibitemShut {NoStop}%
\bibitem [{Not()}]{Note}%
  \BibitemOpen
  \href@noop {} {\ }\bibinfo {note} {Notice that $|101010\rangle_{\rm TFB}$ is not equal to $|101010\rangle_{\rm FCI}$ and the latter one denotes the GS of $\nu=1/2$ FCI.}\BibitemShut {Stop}%
\bibitem [{\citenamefont {Haldane}(2011)}]{Geometry1}%
  \BibitemOpen
  \bibfield  {author} {\bibinfo {author} {\bibfnamefont {F.~D.~M.}\ \bibnamefont {Haldane}},\ }\bibfield  {title} {\bibinfo {title} {Geometrical description of the fractional quantum hall effect},\ }\href {https://doi.org/10.1103/PhysRevLett.107.116801} {\bibfield  {journal} {\bibinfo  {journal} {Phys. Rev. Lett.}\ }\textbf {\bibinfo {volume} {107}},\ \bibinfo {pages} {116801} (\bibinfo {year} {2011})}\BibitemShut {NoStop}%
\bibitem [{\citenamefont {Yang}(2013)}]{Geometry_Yang1}%
  \BibitemOpen
  \bibfield  {author} {\bibinfo {author} {\bibfnamefont {K.}~\bibnamefont {Yang}},\ }\bibfield  {title} {\bibinfo {title} {Geometry of compressible and incompressible quantum hall states: Application to anisotropic composite-fermion liquids},\ }\href {https://doi.org/10.1103/PhysRevB.88.241105} {\bibfield  {journal} {\bibinfo  {journal} {Phys. Rev. B}\ }\textbf {\bibinfo {volume} {88}},\ \bibinfo {pages} {241105} (\bibinfo {year} {2013})}\BibitemShut {NoStop}%
\bibitem [{\citenamefont {Can}\ \emph {et~al.}(2014)\citenamefont {Can}, \citenamefont {Laskin},\ and\ \citenamefont {Wiegmann}}]{Geometry2}%
  \BibitemOpen
  \bibfield  {author} {\bibinfo {author} {\bibfnamefont {T.}~\bibnamefont {Can}}, \bibinfo {author} {\bibfnamefont {M.}~\bibnamefont {Laskin}},\ and\ \bibinfo {author} {\bibfnamefont {P.}~\bibnamefont {Wiegmann}},\ }\bibfield  {title} {\bibinfo {title} {Fractional quantum hall effect in a curved space: Gravitational anomaly and electromagnetic response},\ }\href {https://doi.org/10.1103/PhysRevLett.113.046803} {\bibfield  {journal} {\bibinfo  {journal} {Phys. Rev. Lett.}\ }\textbf {\bibinfo {volume} {113}},\ \bibinfo {pages} {046803} (\bibinfo {year} {2014})}\BibitemShut {NoStop}%
\bibitem [{\citenamefont {Can}\ \emph {et~al.}(2015)\citenamefont {Can}, \citenamefont {Laskin},\ and\ \citenamefont {Wiegmann}}]{Geometry3}%
  \BibitemOpen
  \bibfield  {author} {\bibinfo {author} {\bibfnamefont {T.}~\bibnamefont {Can}}, \bibinfo {author} {\bibfnamefont {M.}~\bibnamefont {Laskin}},\ and\ \bibinfo {author} {\bibfnamefont {P.~B.}\ \bibnamefont {Wiegmann}},\ }\bibfield  {title} {\bibinfo {title} {Geometry of quantum hall states: Gravitational anomaly and transport coefficients},\ }\href {https://doi.org/https://doi.org/10.1016/j.aop.2015.02.013} {\bibfield  {journal} {\bibinfo  {journal} {Annals of Physics}\ }\textbf {\bibinfo {volume} {362}},\ \bibinfo {pages} {752} (\bibinfo {year} {2015})}\BibitemShut {NoStop}%
\bibitem [{\citenamefont {Yang}(2016)}]{Geometry_Yang2}%
  \BibitemOpen
  \bibfield  {author} {\bibinfo {author} {\bibfnamefont {K.}~\bibnamefont {Yang}},\ }\bibfield  {title} {\bibinfo {title} {Acoustic wave absorption as a probe of dynamical geometrical response of fractional quantum hall liquids},\ }\href {https://doi.org/10.1103/PhysRevB.93.161302} {\bibfield  {journal} {\bibinfo  {journal} {Phys. Rev. B}\ }\textbf {\bibinfo {volume} {93}},\ \bibinfo {pages} {161302} (\bibinfo {year} {2016})}\BibitemShut {NoStop}%
\bibitem [{\citenamefont {Can}\ \emph {et~al.}(2016)\citenamefont {Can}, \citenamefont {Chiu}, \citenamefont {Laskin},\ and\ \citenamefont {Wiegmann}}]{Geometry4}%
  \BibitemOpen
  \bibfield  {author} {\bibinfo {author} {\bibfnamefont {T.}~\bibnamefont {Can}}, \bibinfo {author} {\bibfnamefont {Y.~H.}\ \bibnamefont {Chiu}}, \bibinfo {author} {\bibfnamefont {M.}~\bibnamefont {Laskin}},\ and\ \bibinfo {author} {\bibfnamefont {P.}~\bibnamefont {Wiegmann}},\ }\bibfield  {title} {\bibinfo {title} {Emergent conformal symmetry and geometric transport properties of quantum hall states on singular surfaces},\ }\href {https://doi.org/10.1103/PhysRevLett.117.266803} {\bibfield  {journal} {\bibinfo  {journal} {Phys. Rev. Lett.}\ }\textbf {\bibinfo {volume} {117}},\ \bibinfo {pages} {266803} (\bibinfo {year} {2016})}\BibitemShut {NoStop}%
\bibitem [{\citenamefont {Wu}\ \emph {et~al.}(2017)\citenamefont {Wu}, \citenamefont {Tu},\ and\ \citenamefont {Sreejith}}]{Geometry_HHTu}%
  \BibitemOpen
  \bibfield  {author} {\bibinfo {author} {\bibfnamefont {Y.-H.}\ \bibnamefont {Wu}}, \bibinfo {author} {\bibfnamefont {H.-H.}\ \bibnamefont {Tu}},\ and\ \bibinfo {author} {\bibfnamefont {G.~J.}\ \bibnamefont {Sreejith}},\ }\bibfield  {title} {\bibinfo {title} {Fractional quantum hall states of bosons on cones},\ }\href {https://doi.org/10.1103/PhysRevA.96.033622} {\bibfield  {journal} {\bibinfo  {journal} {Phys. Rev. A}\ }\textbf {\bibinfo {volume} {96}},\ \bibinfo {pages} {033622} (\bibinfo {year} {2017})}\BibitemShut {NoStop}%
\bibitem [{\citenamefont {Liou}\ \emph {et~al.}(2019)\citenamefont {Liou}, \citenamefont {Haldane}, \citenamefont {Yang},\ and\ \citenamefont {Rezayi}}]{Geometry5}%
  \BibitemOpen
  \bibfield  {author} {\bibinfo {author} {\bibfnamefont {S.-F.}\ \bibnamefont {Liou}}, \bibinfo {author} {\bibfnamefont {F.~D.~M.}\ \bibnamefont {Haldane}}, \bibinfo {author} {\bibfnamefont {K.}~\bibnamefont {Yang}},\ and\ \bibinfo {author} {\bibfnamefont {E.~H.}\ \bibnamefont {Rezayi}},\ }\bibfield  {title} {\bibinfo {title} {Chiral gravitons in fractional quantum hall liquids},\ }\href {https://doi.org/10.1103/PhysRevLett.123.146801} {\bibfield  {journal} {\bibinfo  {journal} {Phys. Rev. Lett.}\ }\textbf {\bibinfo {volume} {123}},\ \bibinfo {pages} {146801} (\bibinfo {year} {2019})}\BibitemShut {NoStop}%
\end{thebibliography}%
\end{document}